\documentclass[11pt]{iopart}

\usepackage{graphicx}
\usepackage{amssymb}
\usepackage{mathrsfs}
\usepackage{subfigure}
\usepackage{color}
\usepackage{bm}
\usepackage{hyperref}
\usepackage{amsopn}

\DeclareMathOperator{\trace}{tr}
\DeclareMathOperator{\sinc}{sinc}
\DeclareMathOperator{\allsums}{\sum\!\!\int}

\begin{document}

\title{Dynamical quantum correlations of Ising models on an arbitrary
  lattice and their resilience to decoherence}

\author{M Foss-Feig$^{1,}$$^2$, K R A Hazzard$^{3,}$$^4$, J J
  Bollinger$^4$, A M Rey$^{3,}$$^4$, and C W Clark$^{1,}$$^2$}
\address{$^1$National Institute of Standards and Technology, Gaithersburg, MD, 20899, USA}
\address{$^2$Joint Quantum Institute and the Department of Physics, University of Maryland, College Park, MD, 20742, USA}
\address{$^3$JILA and the Department of Physics, University of Colorado,
  Boulder, CO 80309-0440, USA}
\address{$^4$National Institute of Standards and Technology, Boulder,
  Colorado 80305, USA}

\begin{abstract}
Ising models, and the physical systems described by them, play a central role in generating
entangled states for use in quantum metrology and quantum
information.  In particular, ultracold atomic gases, trapped ion
systems, and Rydberg atoms realize long-ranged Ising models, which
even in the absence of a transverse field can give rise to highly
non-classical dynamics and long-range \emph{quantum} correlations.  In the first part of this paper, we present a detailed theoretical framework for studying the dynamics of
such systems driven (at time $t=0$) into arbitrary unentangled non-equilibrium
states, thus greatly extending and unifying the work of Ref. \cite{fossfeig1}.
Specifically, we derive exact expressions for closed-time-path ordered
correlation functions, and use these to study experimentally relevant observables, e.g. Bloch vector and
spin-squeezing dynamics.  In the second part, these correlation
functions are then used to derive closed-form expressions for the
dynamics of arbitrary spin-spin correlation
functions in the presence of both $T_1$ (spontaneous spin relaxation/excitation) and $T_2$
(dephasing) type decoherence processes.  Even though the
decoherence is local, our solution reveals that the competition between
Ising dynamics and $T_1$ decoherence gives rise to an emergent
\emph{non-local} dephasing effect, thereby drastically amplifying the
degradation of quantum correlations.  In addition to identifying
the mechanism of this deleterious effect, our solution points
toward a scheme to eliminate it via measurement-based coherent feedback.
\end{abstract}

\maketitle
\tableofcontents

\section{Introduction \label{sec0}}

Interacting spin models provide a remarkably accurate description
of a diverse set of physical systems, ranging from quantum
magnetic materials \cite{auerbach,lacroix2011,sachdev2008} to quantum dots \cite{hanson}, nitrogen vacancy
centers \cite{prawer}, superconducting qubit arrays \cite{mooij}, ultracold
atomic gases \cite{blochrmp}, and trapped ions \cite{poras,sorensen}.
Despite being relatively simple, and often admitting accurate
theoretical descriptions, they support a variety of complex equilibrium properties found in real materials, e.g. emergent
spatial ordering \cite{auerbach}, quantum criticality \cite{sachdevQPT}, and nontrivial
topological phases \cite{PhysRevB.53.3304,kitaev}.  While the equilibrium physics of the
simplest quantum spin models is, with many notable
exceptions, fairly well understood, the study of driven, dissipative, and otherwise
non-equilibrium behavior is comparatively full of open questions:
Under what circumstances can equilibrium correlations survive coupling to a noisy environment \cite{leggett,PhysRevLett.78.167,dallatorre}?  To what extent do the concepts
of criticality and universality extend to dynamics and non-equilibrium
steady-states \cite{PhysRevLett.97.236808,PhysRevLett.95.177201,diehl2,prosen}?
Can interesting quantum-correlated states be stabilized by (rather than
degraded by) decoherence
\cite{kraus,verstraete,fossfeig,barreiro,PhysRevLett.107.080503,PhysRevLett.109.020403,PhysRevA.85.043620,lee,GopalakrishnanLee}?  In recent years, it has become increasingly apparent that non-equilibrium
dynamics is ideally suited to investigation by quantum simulation \cite{lamacraft_2011}, making such questions especially timely
and important.  Moreover, there are many examples
where interesting non-equilibrium states of matter are \emph{more readily} achievable than low temperature equilibrium
states in ultracold neutral gases \cite{Trotzky18012008}, polar
molecules \cite{hazzard}, and trapped ions \cite{britton}.

With these motivations in mind, in this manuscript we develop a general formalism for
calculating unequal-time correlation functions of arbitrary-range
Ising models driven far out of equilibrium at time $t=0$, thus establishing a comprehensive toolbox for the description of
non-equilibrium dynamics in a simple context.  In addition to
providing a tractable example of quantum many-body spin
dynamics, the Ising model is realized to a good approximation in a variety of experimentally relevant systems.  And, despite its
simplicity, Ising spin dynamics is known to be useful for the production of entangled states with
applications in quantum information and precision metrology
\cite{ueda}.  Our results constitute a unified approach to describing
experiments aimed at producing such states \cite{esteve,gross,leibfried,monz,monroe1,britton}, and
facilitate a quantitative treatment of a variety of unavoidable experimental complications,
e.g. long-range (but not infinite-range) interactions and initial-state imperfections.

Ultracold atomic systems are also well suited to the controlled
inclusion of dissipation, prompting a number of theoretical proposals
to exploit dissipation for the creation of interesting quantum states
\cite{diehl1,diehl2,kraus,verstraete}---remarkably, such ideas are already coming to experimental fruition
\cite{barreiro,krauter}.  Verifying that these experimental systems behave in the expected
manner in the presence of dissipation, however, is extremely challenging, in large part due to the numerical complexity
of simulating dynamics in open quantum systems and the scarcity of
exact solutions.  The Ising model, especially as implemented in
trapped ion experiments \cite{monroe1,monroe2,britton}, poses a unique opportunity to study the
effects of dissipation in a controlled and, as we will show, theoretically tractable
setting.  In the absence of dissipation, an important issue in the
Ising model is whether ground state correlations survive the application of
an equilibrium coherent drive that does not commute with the interactions---i.e. a transverse field.  In the
dissipative Ising model an analogous question can be posed: How does the system respond
to being driven \emph{incoherently} by processes that do not commute
with the Ising interaction?  Our formalism for the calculation of
unequal-time correlation functions allows us to definitively answer this question.

Quite surprisingly, non-equilibrium dynamics in the Ising model remains solvable in the
presence of non-commuting dissipation \cite{fossfeig1}, even for
\emph{completely arbitrary} spatial dependence of the Ising
couplings (and therefore in any dimension).  This manuscript substantially
extends the groundwork laid in Ref. \cite{fossfeig1}, where the
quantum trajectories technique was used to obtain a closed-form
solution of non-equilibrium dynamics for a special class of initial
states.  The present work not only provides a more direct, unified, and comprehensive
exposition of the relevant theory, but also generalizes those results
to include a broader class of initial conditions and observables, and
applies the solutions to a number of experimentally relevant problems (most of which had previously been explored only
by numerical or approximate techniques, if at all).  We also develop a clear physical picture of
the interplay between coherent interactions and spontaneous spin flips, which reveals that $T_1$ decoherence is much more
detrimental to entanglement generation than might be naively
expected.  However, our solution also points toward a
measurement-based feedback scheme that can mitigate its
detrimental effects.

The organization of the manuscript is as follows.  In Sec. \ref{sec1}
we consider the coherent (Hamiltonian) far-from-equilibrium dynamics
of an Ising model with arbitrary spin-spin couplings.  Our results
comprise a unified framework for calculating unequal-time correlation
functions starting from arbitrary unentangled pure states.  As special
cases, these results will be applied to calculating Bloch-vector
dynamics, arbitrary equal time spin-spin correlation functions, and
two-time dynamical correlation functions.  These results are
substantially more general than any already available in the literature \cite{emch,radin,ueda,kastner2,fossfeig1}, and will help quantify the
quantum-enhanced precision in metrology experiments using trapped ions and
ultracold neutral atomic gases.  In Sec. \ref{sec2} we consider the
effect of Markovian decoherence on this dynamics, incorporating
dephasing, spontaneous excitation, and spontaneous relaxation.  Because the
excitation and relaxation processes do not commute with the Ising dynamics,
including them is especially nontrivial: we work in the interaction picture of the Ising Hamiltonian and
incorporate them as time-dependent perturbations.  Terms in the perturbative
expansion are evaluated using the tools laid out in Sec. \ref{sec1},
and by summing the perturbation theory to all orders we obtain an
exact (closed-form) description of the dissipative dynamics of
arbitrary two-point correlation functions.  This is in stark contrast
to the behavior of a \emph{coherently} driven Ising model, where such a
perturbative expansion cannot in general be resummed.  An interesting
feature revealed by our exact solution is that the spin dynamics undergoes an oscillatory-to-damped
transition at a critical dissipation strength, which---in the absence
of a coherent drive---cannot occur at the single-particle or
mean-field level. This feature, along with the more general structure
of our solution, is demonstrated by solving for spin dynamics in
a nearest-neighbor Ising model.  We conclude this section by casting our solution in
terms of a clear physical picture, in which $T_1$
decoherence (spontaneous excitation/relaxation), through its interplay
with the Ising dynamics, gives rise to an emergent non-local dephasing
process.  Section \ref{sec3} applies the solution to calculating
experimentally relevant observables in a dissipative version of the
one-axis twisting model.  We show that the emergent dephasing
discussed in Sec. \ref{sec2} severely diminishes the precision
enhancement achievable compared to that obtained in the absence of decoherence.  However, we
also show that, in special cases, this degradation can be prevented by a
measurement-based feedback mechanism.  In Sec. \ref{sec4} we
summarize our results and pose a number of unanswered
questions that would be interesting to address in future work.

\section{Coherent dynamics in Ising models \label{sec1}}
Our goal is to develop a unified strategy for describing the
dynamics of a collection of spin-$1/2$
particles interacting via Ising couplings and initially (at time $t=0$) driven far
out of equilibrium.  In the absence of a magnetic field, the most
general form for an Ising model is\footnote{Note that many references
  studying long-ranged Ising models---i.e. those for which $\sum_{j}J_{ij}$
is an extensive quantity---often normalize the interaction by dividing
by the number of spins $\mathcal{N}$.  We drop this constant here to avoid
cluttering the notation.}
\begin{equation}
\label{modelham}
\mathcal{H}=\sum_{j<k}J_{jk}\hat{\sigma}_j^z\hat{\sigma}_{k}^{z},
\end{equation}
where $\hat{\sigma}_j^z$ are $z$ Pauli matrices and the indices $j,k$
label lattice sites located at spatial positions $\bm{r}_j$.  The coupling constants $J_{jk}$ are left completely arbitrary, and
hence there is not necessarily any notion of dimensionality.  In many
physical realizations of this Hamiltonian, such as trapped ions,
neutral atoms, or Rydberg atoms, the couplings exhibit a roughly
power-law spatial dependence, $J_{jk}=J|\bm{r}_j-\bm{r}_k|^{-\zeta}$.

Because there is no transverse field (a term $\propto h\sum_{j}\hat{\sigma}^x_j$), the eigenstates of $\mathcal{H}$ can
always be chosen to be simultaneous eigenstates of all the $\hat{\sigma}^z_j$ (with eigenvalues $\sigma_j^z=\pm1$).
As a result, the partition function $Z(\beta)=\trace\left[e^{-\beta\mathcal{H}}\right]$ (with $\beta$ the
inverse temperature) is identical to that of a classical Ising model, and the equivalence of all
equilibrium properties follows.  This is the sense in which the Ising
model without a transverse field is often said to be ``classical''
(even though it is a quantum Hamiltonian acting on vectors in a
Hilbert space).  In passing we note that classical Ising
models can, of course, be highly nontrivial: for example, disordered or frustrated couplings give rise to classical glassiness
\cite{PhysRevLett.35.1792,PhysRevLett.23.17,PhysRevLett.54.1321}.  Out of equilibrium, however, this notion of classicality is inapplicable.  While a
thermal density matrix $\rho(\beta)=Z^{-1}e^{-\beta\mathcal{H}}$ commutes with $\mathcal{H}$, the density matrix
describing some non-equilibrium initial conditions will \emph{not} in general commute
with the Hamiltonian, and nontrivial dynamics will ensue.  This
dynamics---which has no direct analogue in the \emph{classical} Ising model---is generically
characterized by the growth of entanglement, leading in some special cases to spin-states with
applications in quantum information and precision metrology\cite{ueda}.

\begin{figure}[h!]
\centering
\subfiguretopcaptrue
\subfigure[][]{
\label{Fig1a}
\includegraphics[width=10.6cm]{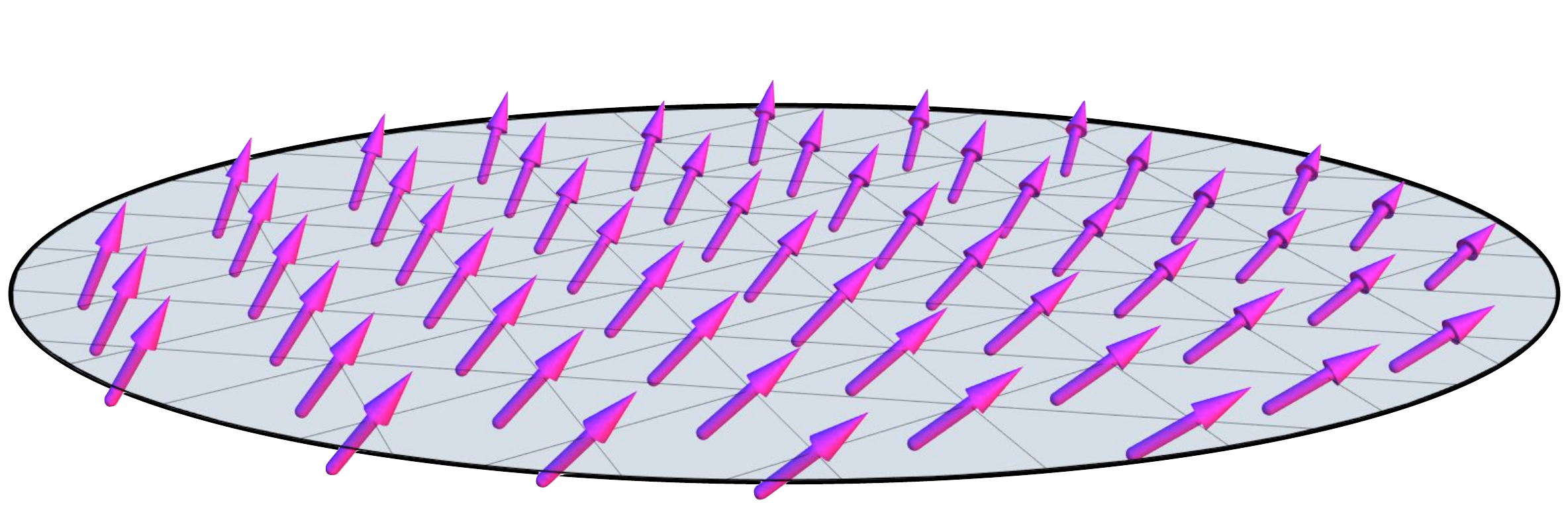}
}
\subfigure[][]{
\label{Fig1b}
\includegraphics[width=3.8cm]{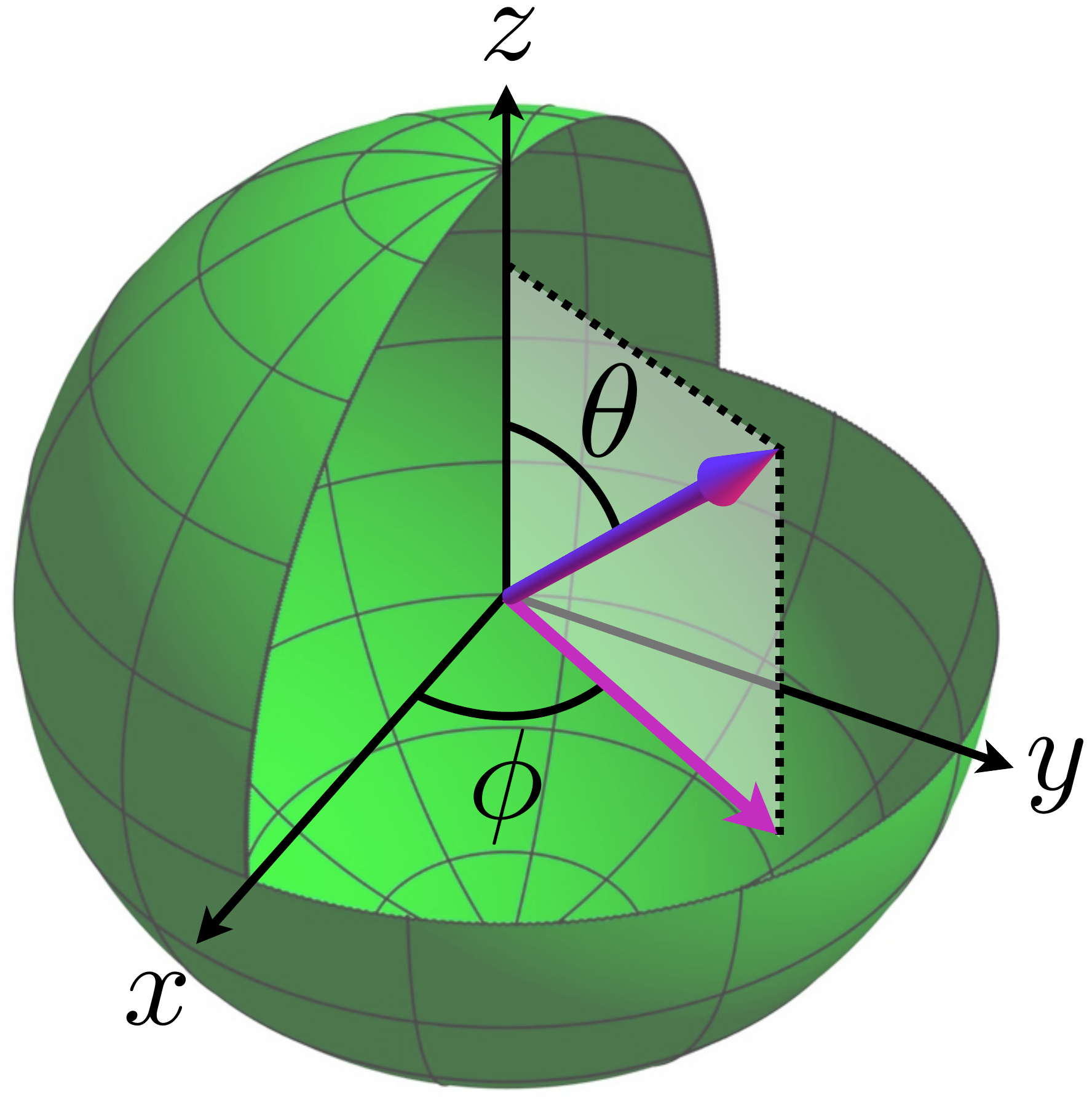}
}
\caption{The only restriction on the initial
state is that it be unentangled (i.e. a product state over the various
sites in the spin model).  For example, one can imagine \subref{Fig1a} an initial
state with some slow spatial variations in the spin angles due to
inhomogeneities in the pulse strength of a Ramsey-type experiment.
The notation used to characterize the state of any one spin is shown
on the Bloch sphere in \subref{Fig1b}.}
\label{fig2}
\end{figure}

Everywhere in this manuscript, we assume the system starts in a pure state that
is a direct product between the various spins [Fig. \ref{Fig1a}].\footnote{Unentangled but mixed
initial density matrices can be easily accounted for by suitable
averaging of the expressions given over the initial ensemble.}  The most general such
state can be specified by choosing spherical angles $\theta_j$ and
$\phi_j$ describing the orientation of the spin at each site $j$
[Fig. \ref{Fig1b}].  Defining
\begin{equation}
f_{j}(1)=e^{-i\phi_j/2}\cos\frac{\theta_j}{2},~~~~~f_{j}(-1)=e^{i\phi_j/2}\sin\frac{\theta_j}{2},
\end{equation}
and states $|\sigma_j\rangle$ that are eigenstates of $\hat{\sigma}_j^z$ with
eigenvalues $\sigma_j=\pm 1$, such a state can be written
\begin{eqnarray}
|\Psi(t=0)\rangle&=&\bigotimes_{j}\left|\psi_j\right\rangle\\
&=&\bigotimes_{j}\sum_{\sigma_{j}}f_{j}(\sigma_j)|\sigma_{j}\rangle.
\end{eqnarray}
For uniform $\theta_j=\theta$ and $\phi_j=\phi$, the state $|\Psi(0)\rangle$ is
frequently encountered in experiments implementing Ramsey
spectroscopy \cite{PhysRevA.46.R6797,britton,hazzard}, and spatially varying angles could be
used, for example, to describe the effects of defects or excitation
inhomogeneities in such experiments \cite{Campbell17042009,PhysRevLett.103.260402,swallows}.

Essentially all properties of the non-equilibrium dynamics are contained in
unequal-time correlation functions of the spin
operators $\hat{\sigma}_j^{\pm}$ and $\hat{\sigma}_{j}^z$ (these
subsume, of course, the time evolution of all equal-time
correlation functions).  We focus first on the case where only
operators $\hat{\sigma}_j^{\pm}$ occur
\begin{equation}
\label{correlations}
\mathcal{G}=\left\langle\mathcal{T}_{\mathcal{C}}\left(\hat{\sigma}^{a_n}_{j_n}(t_n^{*})\dots\hat{\sigma}^{a_1}_{j_1}(t_1^{*})\hat{\sigma}^{b_m}_{k_m}(t_m)\dots\hat{\sigma}^{b_1}_{k_1}(t_1)\right)\right\rangle.
\end{equation}
Here $a,b=\pm$, and the time dependence of the operators is given in the Heisenberg
picture of $\mathcal{H}$
\begin{equation}
\hat{\sigma}_j^a(t)=e^{it\mathcal{H}}\hat{\sigma}_j^ae^{-it\mathcal{H}}.
\end{equation}
The time-ordering operator $\mathcal{T}_{\mathcal{C}}$ orders all
operators along a closed-time-path $\mathcal{C}$ shown in  Fig. \ref{fig2}, with times $t$
occurring on the forward path and times $t^{*}$ occurring along the
backwards path.  This closed-time path ordering occurs naturally, for instance,
in any perturbative treatment of additional non-commuting terms in the
Hamiltonian.  In Sec. \ref{sec2} we encounter this situation when
treating a dissipative coupling to an environment, but the same
structure occurs for coherent couplings, e.g. a transverse field $h\sum_j\hat{\sigma}^x_j$.  Our goal in what
follows is to obtain analytic expressions for such correlation functions in full generality,
and then apply them to calculating a variety of experimentally
relevant quantities.
\begin{figure}[h!]
\centering
\includegraphics[width=11cm]{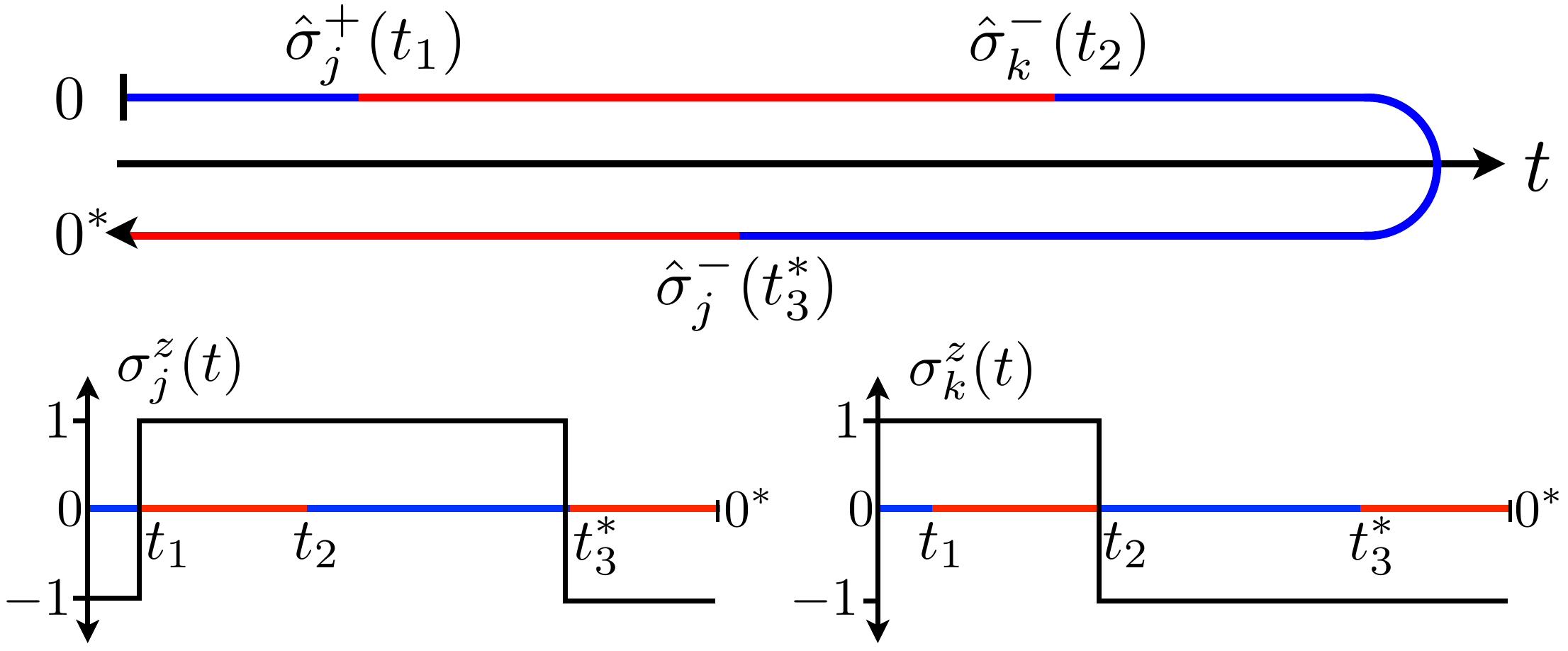}
\caption{Graphical representation of a sample correlation function
  $\mathcal{G}=\left\langle\mathcal{T}_{\mathcal{C}}\left(\hat{\sigma}_{j}^{-}(t_3^{*})\hat{\sigma}_{k}^{-}(t_2)\hat{\sigma}_j^{+}(t_1)\right)\right\rangle$.
  Here $\alpha_j=\alpha_k=0$, $\alpha_{l\neq j,k}=1$, and the time-dependent functions $\sigma_{j}^{z}(t)$ and $\sigma_{k}^{z}(t)$ are shown in
  the bottom two panels.}
\label{fig2}
\end{figure}
In order to describe $\mathcal{G}$ concisely, it is useful to define a variable $\alpha_j$ on each site such that
$\alpha_j=1$ if there are no occurrences of the operators
$\hat{\sigma}_{j}^{a}$ in $\mathcal{G}$, and $\alpha_j=0$ otherwise.
Now we recognize that if an operator $\hat{\sigma}_j^{a=\pm}$ occurs
in $\mathcal{G}$ one or more times, the operator $\hat{\sigma}_j^z$
(appearing in the time evolution operator) is forced to take on a well
defined value $\sigma_j^z(t)$ at \emph{all points in
  time} (see Fig. \ref{fig2}).  As a result, we can rewrite
the correlation function $\mathcal{G}$ as
\begin{equation}
\label{g1}
\mathcal{G}=\left\langle \exp\left(-i\int_{\mathcal{C}} dt\;\mathcal{K}(t)\right)\right\rangle\prod_{j}\left[p_j+\alpha_j\right],
\end{equation}
where
\begin{equation}
p_j=(1-\alpha_j)\bar{f}_j\left[\sigma_j^{z}(0)\right]f_j\left[\sigma_j^z(0^*)\right],
\end{equation}
($t=0^*$ marking the end of the backwards trajectory,
Fig. \ref{fig2}), $\bar{f}$ is the complex conjugate of $f$, $\int_{\mathcal{C}}$ is a time integral that
runs along the closed-time path, and
\begin{equation}
\fl\mathcal{K}(t)=\frac{1}{2}\sum_{j,k}J_{jk}\left[ (1-\alpha_j)(1-\alpha_k)\sigma^z_j(t)\sigma^z_k(t)+2(1-\alpha_j)\alpha_k\sigma^z_j(t)\hat{\sigma}^z_k+\alpha_j\alpha_k\hat{\sigma}_j^z\hat{\sigma}_k^z\right].
\end{equation}
The first thing to notice is that, since $\int_{\mathcal{C}} dt=0$, the final time-independent term in
$\mathcal{K}$ vanishes. Since this is the \emph{only} non separable
term in $\mathcal{K}$, the remaining time evolution is straightforward
to compute.  It is helpful to define the following parameters that depend on the functions $\sigma_j^z(t)$
\begin{eqnarray}
\varphi_k&\equiv&\sum_{j,k} J_{jk}(1-\alpha_{j})\alpha_k\int_{\mathcal{C}}
dt\;\sigma^{z}_{j}(t)\\
\vartheta&\equiv&\frac{1}{2}\sum_{j,k}J_{jk}(1-\alpha_{j}) (1-\alpha_{k})\int_{\mathcal{C}} dt\;\sigma_j^z(t)\sigma_{k}^{z}(t),
\end{eqnarray}
in terms of which
\begin{equation}
\int_{\mathcal{C}}dt\;\mathcal{K}(t)=\vartheta+\sum_k\varphi_k\hat{\sigma}_k^z.
\end{equation}
The time-ordered correlation functions of Eq. (\ref{g1}) can
now be compactly written
\begin{eqnarray}
\label{TOCF}
\mathcal{G}&=&e^{-i\vartheta}\prod_{j}\left\langle\psi_j\right| e^{-i\varphi_j\hat{\sigma}_j^z}\left|\psi_j\right\rangle\prod_j\left(p_j+\alpha_j\right)\\
\label{TOCF2}
&=&e^{-i\vartheta}\prod_{j}\left[p_j+g^{+}_j(\varphi_j)\right],
\end{eqnarray}
with
\begin{equation}
g^{\pm}_j(\varphi)=\alpha_j\left(|f_{j}(1)|^2e^{-i\varphi}\pm
|f_{j}(-1)|^2e^{i\varphi}\right)
\end{equation}
($g^{-}_{j}$ will be useful momentarily).  The equivalence between Eqs. (\ref{TOCF}) and (\ref{TOCF2}) can be understood by explicitly comparing the
expressions inside the product for the two possible situations
$\alpha_j=0,1$: If $\alpha_j=0$, then $\varphi_j=0$ and
the expectation value is unity, whereas when $\alpha_j=1$ we find
$p_j=0$ and the expectation value gives $g^{+}(\varphi_j)$.

The insertion of an operator $\hat{\sigma}_j^z(t)$ inside a
correlation function $\mathcal{G}$, which we denote by writing
$\mathcal{G}\rightarrow\mathcal{G}_j^z$, is relatively straightforward.  If $\alpha_j=0$,
then clearly the substitution $\hat{\sigma}_j^z\rightarrow\sigma_j^z(t)$ does the trick.  If
$\alpha_j=1$, $\hat{\sigma}^z_j$ can be inserted by
recognizing that the variable $\varphi_j$ couples to
$\hat{\sigma}_j^z$ as a source term, and thus the
insertion of $\hat{\sigma}_j^z(t)$ is equivalent to applying
$i\frac{\partial}{\partial\varphi_j}$ to $\mathcal{G}$.  Both
possibilities are captured by writing
\begin{equation}
\label{GF}
\mathcal{G}_j^z=\left((1-\alpha_j)\sigma_j^z(t)+\alpha_j i\frac{\partial}{\partial \varphi_j}\right)\mathcal{G},
\end{equation}
which, using $\partial g^{+}(\varphi)/\partial \varphi=-ig^-(\varphi)$, can be simplified as
\begin{equation}
\mathcal{G}_j^z=e^{-i\vartheta}\left[p_j\sigma^z_j(t)+g_{j}^{-}(\varphi_j)\right]\prod_{k\neq j}\left[p_k+g^+_k\left(\varphi_k\right)\right].
\end{equation}
Notice that if all operators occur at the same time $t$,
e.g. when calculating equal-time correlation functions, then $\vartheta=0$
and $\varphi_j=\sum_{k}J_{jk}(1-\alpha_k) \alpha_j \left(\pm2t\right)$ (with the $\pm$
depending on whether $\hat{\sigma}_k^{\pm}$ is applied to the spin on
site $k$).

\subsection{Bloch vector dynamics}

\begin{figure}[t!]
\centering
\includegraphics[width=15cm]{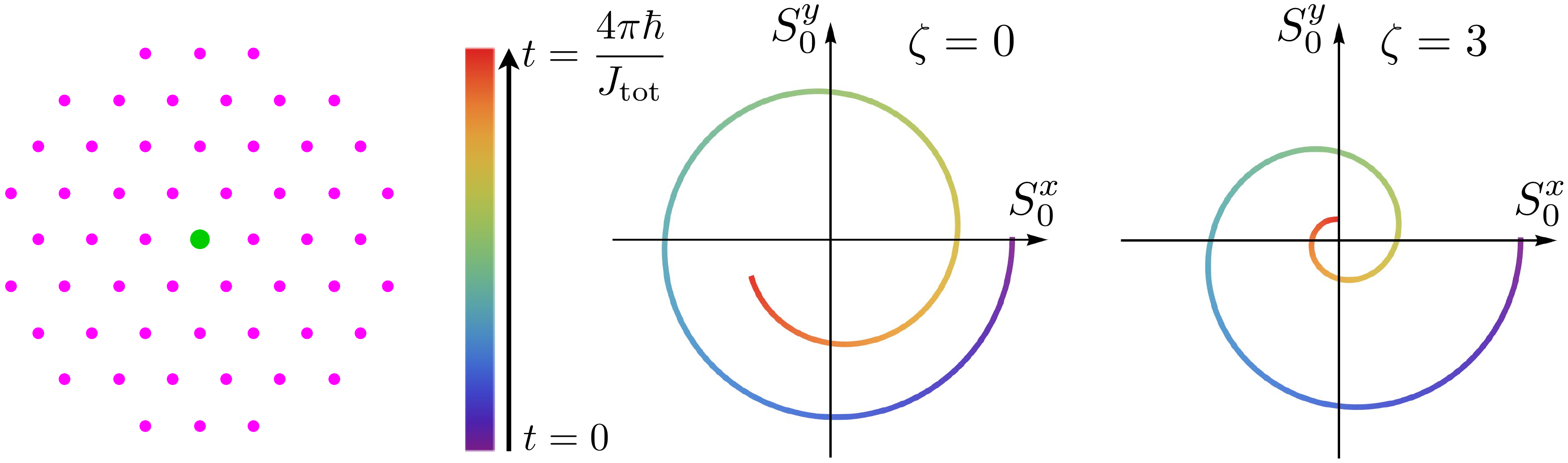}
\caption{Trajectories of the Bloch vector $\bm{S}_0$ projected into the $xy$
  plane, for all-to-all (left) and dipolar (right) couplings.  Here
  $j=0$ labels the central site (green dot in left pannel) of a 55-site
  triangular lattice, and all spins are initialized at $\{\theta_j,\phi_j\}=\{\pi/4,0\}$.  In both cases we choose a nearest neighbor
  coupling $J$, and scale the time by $J_{\mathrm{tot}}=\sum_{j\neq
    0}J/(\bm{r}_j-\bm{r}_0)^{\zeta}$.  This rescaling of time is used
  so that different range interactions give rise to comparable
  precession rates.  Note that at mean-field level the trajectory would close
  on itself.  The inward spiral indicates the growth of quantum correlations and resultant decay of
  the spin length, and---in these rescaled time units---is more significant for
  shorter-range interactions.}
\label{fig3}
\end{figure}

It is now straightforward to calculate the dynamics of the Bloch vectors
\begin{equation}
\bm{S}_j(t)=\frac{1}{2}\{\langle\sigma_{j}^{x}(t)\rangle,\langle\sigma_{j}^{y}(t)\rangle,\langle\hat{\sigma}_j^z(t)\rangle\}.
\end{equation}
Because $\hat{\sigma}^{z}_j$ commutes with the Ising interaction the $z$ component of spin is time independent, and given by
$S_j^z=\frac{1}{2}g^{-}_j(0)=\frac{1}{2}\cos\theta_j$.  The transverse spin components $S_j^x(t)$ and
$S_j^y(t)$ can be obtained from the real and imaginary parts, respectively, of
$\langle\hat{\sigma}^{+}_j(t)\rangle$.  A straightforward application
of  Eq. (\ref{TOCF}) gives
\begin{eqnarray}
\label{sigmaplus}
\langle\hat{\sigma}^{+}_j(t)\rangle&=&\bar{f}_j(1)f^{\phantom *}_j(-1)\prod_{k\neq
j}g^{+}_{k}(2J_{jk} t)\\
&=&\frac{1}{2}e^{i\phi_j}\sin\theta_j\prod_{k\neq
  j}\left(\cos 2J_{jk}t-i\sin 2J_{jk}t\cos\theta_k\right). \nonumber
\end{eqnarray}
For the special case where all spins point along $\theta=\pi/2$ at
$t=0$ there is pure decay of the Bloch vector without any rotation.
In Fig. (\ref{fig3}) we show the projection into the $xy$ plane of the Bloch vector
$\bm{S}_0(t)$ (where $j=0$ labels the central site of a 55-site triangular lattice) for an
initial state in which all spins point in a single direction lying outside of the $xy$ plane ($\theta=\pi/4$, $\phi=0$).  The
Bloch vector spirals inwards: The precession can be understood as a mean-field
effect \cite{britton}, with the spin rotating due to the average magnetization of
the other spins, while the decay is due to the development of quantum
correlations.  Note that, in a finite system, the length $S(t)=|\bm{S}(t)|$ of the \emph{total}
Bloch vector $\bm{S}(t)=\sum_j\bm{S}_j(t)$ decays even at the mean-field
level for any $\zeta\neq0$.  This decay, however, is due to the
existence of a spatially inhomogeneous mean-field, and cannot be
associated with the development of correlations.

\subsection{Equal time correlation functions\label{coherentsqueezing}}

Spin-spin correlation functions can be calculated just as easily from
Eqs. (\ref{TOCF}) and (\ref{GF}).  All two-point correlation functions can be
calculated from the four quantities
\begin{eqnarray}
\mathcal{C}^{zz}_{jk}(t)&\equiv&\langle
\hat{\sigma}_{j}^{z}(t)\hat{\sigma}_k^{z}(t)\rangle\\
\mathcal{C}^{+z}_{jk}(t)&\equiv&\langle
\hat{\sigma}_{j}^{+}(t)\hat{\sigma}_k^{z}(t)\rangle\\
\mathcal{C}^{++}_{jk}(t)&\equiv&\langle
\hat{\sigma}_{j}^{+}(t)\hat{\sigma}_k^{+}(t)\rangle\\
\mathcal{C}^{+-}_{jk}(t)&\equiv&\langle
\hat{\sigma}_{j}^{+}(t)\hat{\sigma}_k^{-}(t)\rangle
\end{eqnarray}
and their complex conjugates.  Since the Hamiltonian commutes with all
$\hat{\sigma}_j^z$, the first one is given trivially by
$\mathcal{C}^{zz}_{jk}=g^-_j(0)g^-_k(0)=\cos\theta_j\cos\theta_k$.  The second one can be
obtained from Eqs. (\ref{GF}) and (\ref{sigmaplus}) as
\begin{eqnarray}
\mathcal{C}_{jk}^{+z}&=&\bar{f}_j(1)f_j(-1)g^{-}_{k}(2J_{jk}t)\prod_{l\neq
j,k}g^{+}_{l}(2J_{jl} t)\nonumber\\
&=&\frac{1}{2}e^{i\phi_j}\sin\theta_jg^{-}_{k}(2J_{jk}t)\prod_{l\neq
j,k}g^{+}_{l}(2J_{jl} t),
\end{eqnarray}
and the third and fourth are
\begin{eqnarray}
\mathcal{C}_{jk}^{++}&=&\frac{1}{4}e^{i(\phi_j+\phi_k)}\sin\theta_j\sin\theta_k\prod_{l\neq k,j}g^{+}_{l}(2J_{jl}t+2J_{kl}t)\nonumber\\
\mathcal{C}_{jk}^{+-}&=&\frac{1}{4}e^{i(\phi_j-\phi_k)}\sin\theta_j\sin\theta_k\prod_{l\neq k,j}g^{+}_{l}(2J_{jl}-2J_{kl}t)\nonumber
\end{eqnarray}

\begin{figure}[t!]
\centering
\includegraphics[width=8cm]{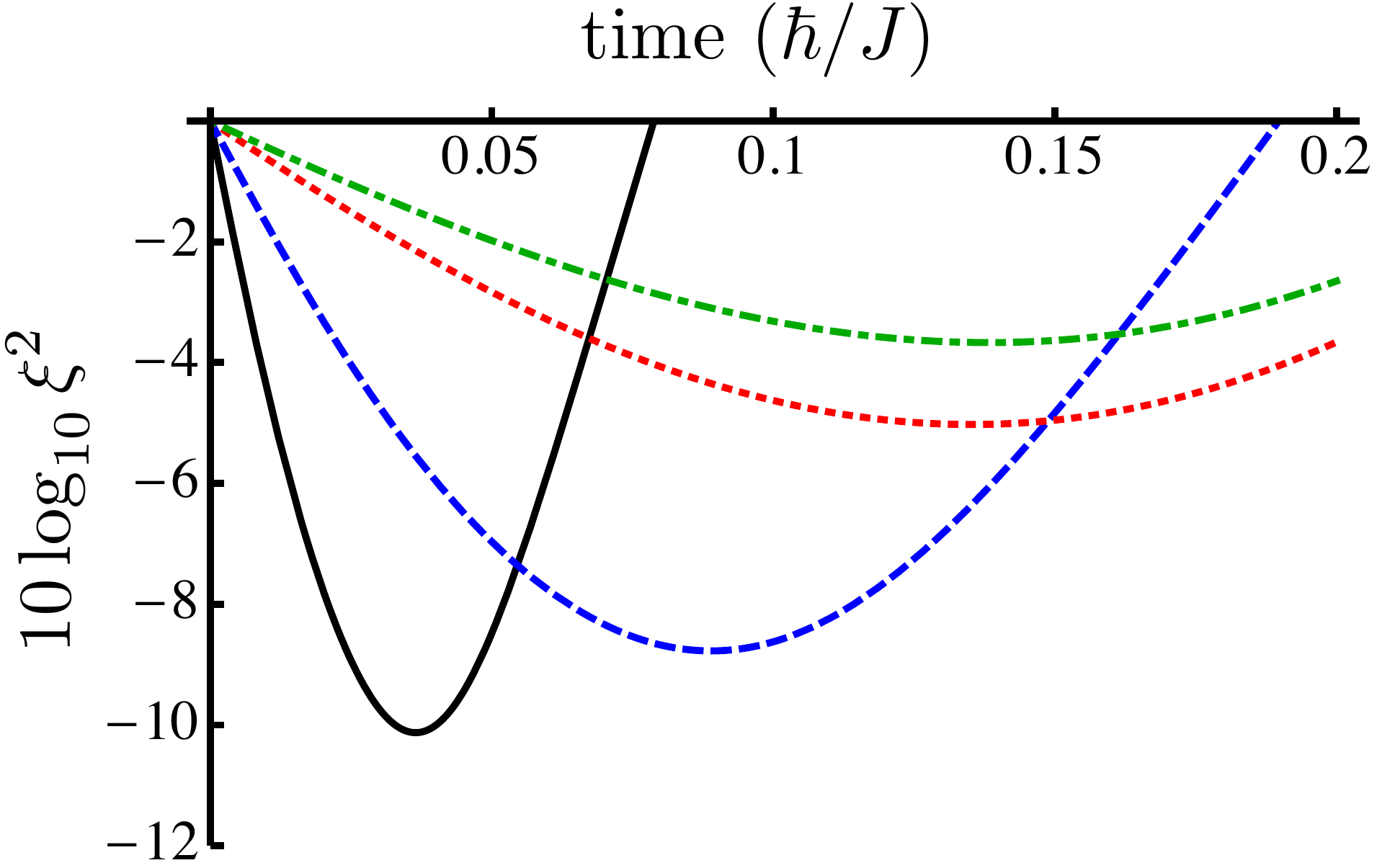}
\caption{Optimal spin squeezing $\xi$ (obtained be minimizing $\xi(t)$ over time) for a variety of power-law couplings
  $J_{ij}=J/|\bm{r}_i-\bm{r}_j|^{\zeta}$.  The different
  curves correspond to:
  infinite ranged ($\zeta=0$, black solid line), coulombic ($\zeta=1$,
  blue dashed line), dipolar ($\zeta=3$, red dotted line), and nearest
  neighbor ($\zeta=\infty$, green dot-dashed line).  For this calculation we
take the spins to all point along the $x$-axis at $t=0$, and use 55
sites of a triangular lattice (the same as shown in Fig. \ref{fig2}).}
\label{FigSqueezing}
\end{figure}

These correlation functions can be used, for example, to calculate the
time dependence of the spin squeezing parameter
\begin{equation}
\xi(t)=\sqrt{\mathcal{N}}\frac{\Delta S_{\mathrm{min}}(t)}{S(t)}.
\end{equation}
Here $\Delta S_{\mathrm{min}}(t)$ is defined to be the
minimum uncertainty along a direction perpendicular to
$\bm{S}(t)$\footnote{If we choose our $x$-axis to be in the direction
of $\bm{S}(t)$, and define
$\hat{S}_{\psi}=\frac{1}{2}\sum_{j}\left(\cos(\psi)\hat{\sigma}^z_j+\sin(\psi)\hat{\sigma}_j^y\right)$,
then $\Delta S_{\mathrm{min}}(t)$ is obtained by minimizing
$(\langle\hat{S}_{\psi}^2\rangle-\langle\hat{S}_{\psi}\rangle^2)^{1/2}$ over $\psi$.}.  The
squeezing parameter determines the phase sensitivity in a suitably
performed Ramsey experiment, which is enhanced over the standard
quantum limit whenever $\xi<1$ \cite{ueda}.  For $\zeta=0$
(infinite-range interactions) the calculation was first performed in
\cite{ueda}.  However, in many experimentally relevant situations the
interactions have some finite range and the maximum achievable
squeezing is diminished (Fig. \ref{FigSqueezing}).

\subsection{Unequal-time correlation functions\label{SecUnequalTime}}

\begin{figure}[t!]
\centering
\subfiguretopcaptrue
\subfigure[][]{
\label{Fig5a}
\includegraphics[width=7cm]{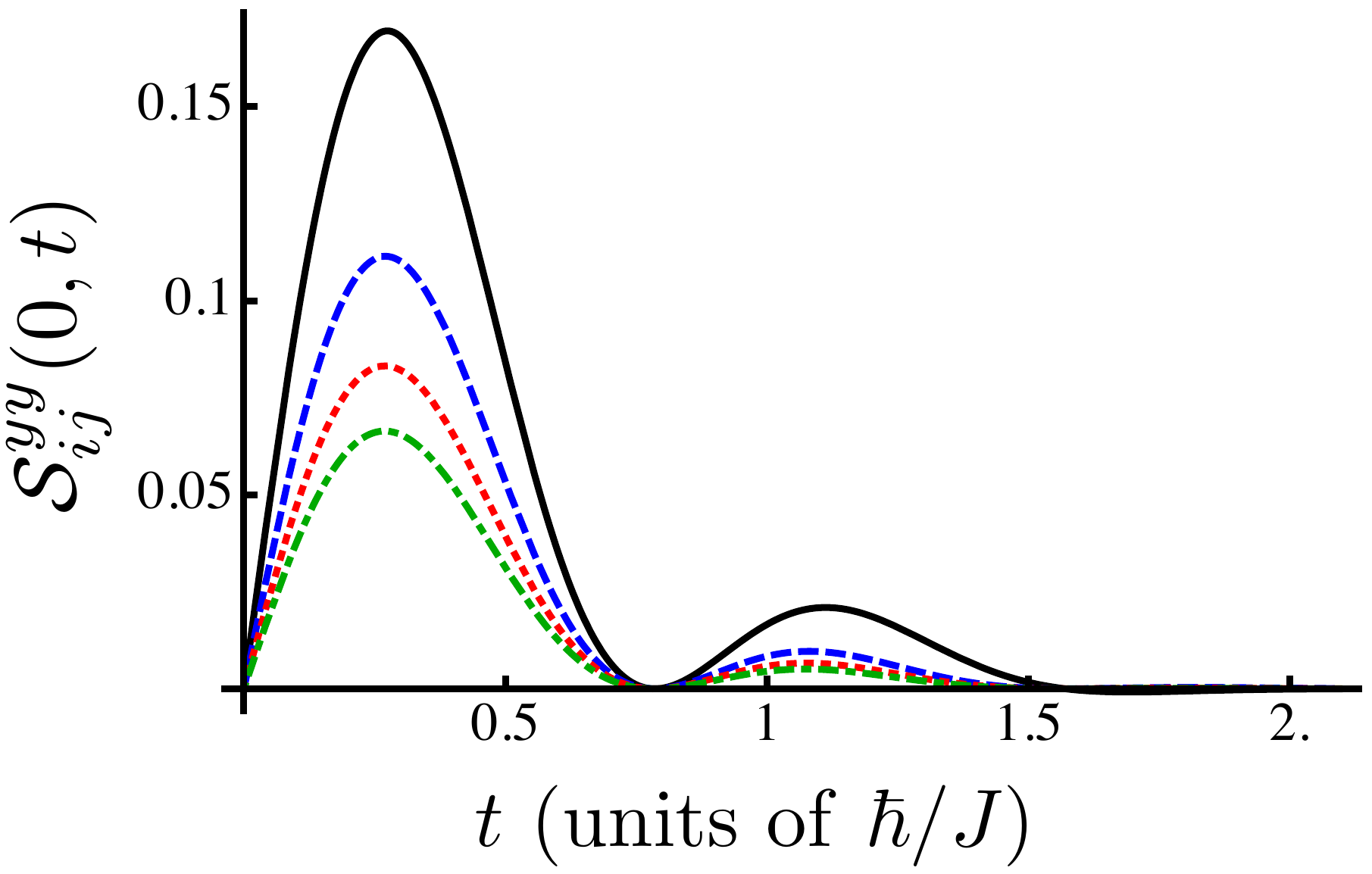}
}
\subfigure[][]{
\label{Fig5b}
\includegraphics[width=5.6cm]{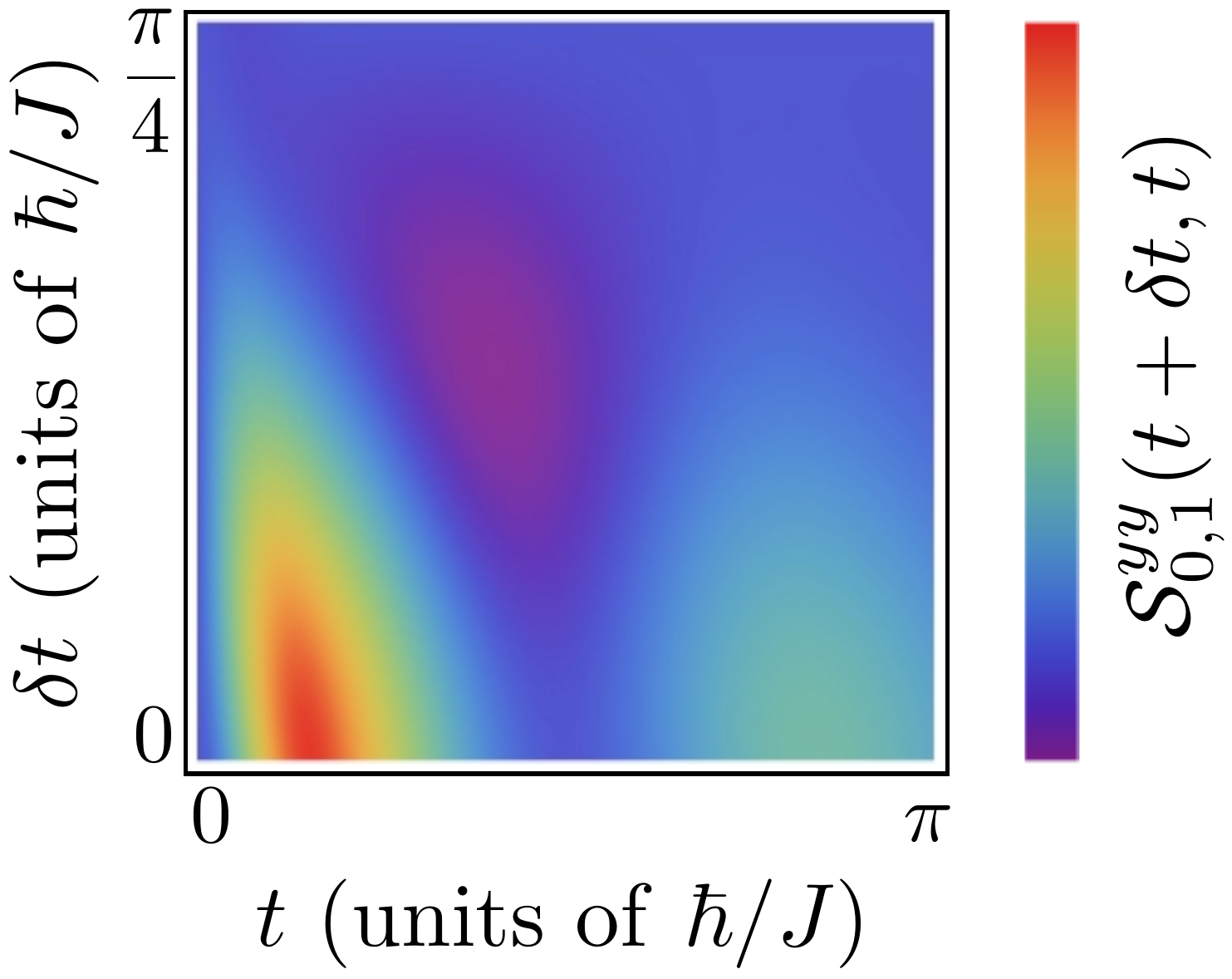}
}
\caption{Connected dynamical correlation function
  $\mathcal{S}_{ij}^{yy}(t_1,t_2)=S_{ij}^{yy}(t_1,t_2)-S_i^y(t_1)S_{j}^{y}(t_2)$
  for a 100 site 1D chain with $\zeta=1$ and nearest-neighbor coupling $J$.  In
  (a) we plot $\mathcal{S}_{i,i+r}^{yy}(0,t)$ for $i=50$ and $r=\{2,3,4,5\}$ (from top to
  bottom).  In (b) we plot $\mathcal{S}_{i,i+1}^{yy}(t,t+\delta t)$
  (again for $i=50$) as a function
of $t$ and $\delta t$.  In both plots the initial state consists of
all spins pointing in the $+x$ direction ($\theta=\pi/2$ and $\phi=0$).}
\label{UnequalTime}
\end{figure}

It is also possible to calculate correlation functions involving
the application of spin operators at different times, which describe the propagation in time of a perturbation to the system.  As an
example, we can easily calculate dynamical response functions of
the form
\begin{eqnarray}
S^{ab}_{ij}(t_1,t_2)&=&\frac{1}{4}\left\langle\hat{\sigma}^a_i(t_1)\hat{\sigma}^b_j(t_2)\right\rangle\nonumber\\
&=&\frac{1}{4}\sin\theta_i\sin\theta_je^{-2iabJ_{ij}(t_2-t_1)}\prod_{k\neq
  i,j}g^{+}\left(2a t_1 J_{ik}+2b t_2 J_{jk}\right).
\end{eqnarray}
These can be combined to calculate dynamical response functions
involving arbitrary Pauli matrices, some examples of which are shown
in Fig. \ref{UnequalTime}.

As we will see in Sec. \ref{sec2}, such dynamical correlation
functions allow us to calculate the effect of spontaneous relaxation and
excitation on the dynamics of the system.  For instance, for all spins
initially polarized along the $x$ axis, the effect of the sudden relaxation of a
spin on site $k$ at time $t_{\mathrm{s}}$ on the time
dependence of $\hat{\sigma}^{x}_j$ (for $j\neq k$) is given by
\begin{eqnarray}
\left\langle\hat{\sigma}^{+}_k(t^{*}_{\mathrm{s}})\hat{\sigma}^{x}_j(t)\hat{\sigma}^{-}_k(t_{\mathrm{s}})\right\rangle&=&\mathrm{Re}\left[e^{2iJ_{jk}(t-2t_{\mathrm{s}})}\prod_{l\neq
  j,k}\cos(2J_{jl}t)\right]\nonumber\\
&=&\cos(2J_{jk}t[1-2t_{\mathrm{s}}/t])\left\langle\hat{\sigma}^{x}_j(t)\right\rangle.
\end{eqnarray}
Note that this is the same time evolution we would obtain if no spontaneous
relaxation had occurred, the $k$'th spin were simply absent, and the $j$'th
spin were coupled to a longitudinal magnetic field of strength
$2J_{jk}(2t_s/t-1)$.  The reason for this behavior is straightforward
when considering the time evolution in the Schr\"odinger picture.  The
application of $\hat{\sigma}^{-}_k$ to the wave function at time
$t_{\mathrm{s}}$ not only forces the $k$'th spin to point down between
times $t_{\mathrm{s}}$ and $t$, but also destroys the piece of the
initial wave function having weight into states with $\sigma_k=-1$.
Hence it is as if the $k$'th spin pointed up for a time $t_{\mathrm{up}}=t_\mathrm{s}$,
and down for a time $t_{\mathrm{down}}=t-t_{\mathrm{s}}$, thus
contributing an inhomogeneous longitudinal magnetic field of strength
$2J_{jk}(t_{\mathrm{up}}-t_{\mathrm{down}})/t=2J_{jk}(2t_{\mathrm{s}}/t-1)$ (the factor of $2$ arises
because the $J_{jk}$ couple to Pauli matrices rather than the
spin-$\frac{1}{2}$ matrices).

\section{Inclusion of dissipation \label{sec2}}

In the previous sections we have treated our Ising spins as a closed
system.  That is, we neglected any coupling that might exist
between our system and the outside world, and thus initially pure
states remained pure throughout the dynamics.  In any physical
realization of the Ising model, this is clearly an idealization; decoherence occurs and often must be accounted for.  For example, Rydberg
atoms suffer from spontaneous emission \cite{Gallagher1994}, while ions couple strongly to fluctuating
classical (electric and magnetic) fields and can decohere through
off-resonant light scattering from the spin-dependent optical dipole
forces used to engineer the Ising interactions (this off-resonant
light scattering can produce spontaneous excitation/relaxation and
dephasing) \cite{uys}.  One way to envision dynamics in an open system is by considering the probabilistic
occurrence of sudden perturbations of the system---quantum jumps---due
to the system-environment coupling \cite{QuantumTrajectories}.  As suggested in
Sec. \ref{SecUnequalTime}, and as will be explained in detail below,
the strategy we have developed for computing unequal-time correlation functions is well suited to describing such effects.

\subsection{Description of the problem}

Given a density matrix $\varrho$ describing a system coupled to a
reservoir, it is always possible to express the expectation value of a
system operator $\hat{\mathcal{A}}$ in terms of the system reduced density matrix
$\mathcal{\rho}=\trace_{\mathrm{R}}\left[\varrho\right]$ as $\langle
\hat{\mathcal{A}}\rangle=\trace_{\mathrm{S}}[\rho\hat{\mathcal{A}}]$.  Here
$\trace_{\mathrm{R}(\mathrm{S})}$ denotes a trace over the reservoir
(system) degrees of freedom---we will drop these subscripts from now
on, since all future instances of $\trace$ refer to a trace over
system degrees of freedom only.  In this language, the effect of a finite system-reservoir coupling is that an
initially pure system density matrix $\rho(0)\equiv|\Psi(0)\rangle\langle\Psi(0)|$ will evolve into a
mixed state (reflecting entanglement between the system and reservoir
degrees of freedom).  When the system-environment coupling is weak (i.e. small compared to
the inverse of relevant system time scales) and the reservoir
correlation time is small, the Born-Markov approximation is justified
and the reduced system density matrix obeys a Markovian master
equation of Lindblad form \cite{gardiner}.  We choose a very general master
equation appropriate for describing the various types of decoherence
relevant to trapped ions \cite{uys}, Rydberg
atoms \cite{Gallagher1994}, and condensed matter systems such as
quantum dots \cite{hanson} and nitrogen vacancy centers \cite{prawer}:
\begin{equation}
\label{meqformal}
\dot{\rho}=-i\mathscr{H}(\rho)-\mathscr{L}_{\mathrm{ud}}(\rho)-\mathscr{L}_{\mathrm{du}}(\rho)-\mathscr{L}_{\mathrm{el}}(\rho),
\end{equation}
where
\begin{eqnarray}
\mathscr{H}(\rho)&=&\left[\mathcal{H},\rho\right]\\
\mathscr{L}_{\mathrm{ud}}(\rho)&=&\frac{\Gamma_{\mathrm{ud}}}{2}\sum_{j}\left(\hat{\sigma}^+_j\hat{\sigma}^-_j\rho+\rho \hat{\sigma}^+_j\hat{\sigma}^-_j-2\hat{\sigma}^{-}_j\rho\hat{\sigma}^+_j\right)\\
\mathscr{L}_{\mathrm{du}}(\rho)&=&\frac{\Gamma_{\mathrm{du}}}{2}\sum_{j}\left(\hat{\sigma}^-_j\hat{\sigma}^+_j\rho+\rho \hat{\sigma}^-_j\hat{\sigma}^+_j-2\hat{\sigma}^{+}_j\rho\hat{\sigma}^-_j\right)\\
\mathscr{L}_{\mathrm{el}}(\rho)&=&\frac{\Gamma_{\mathrm{el}}}{8}\sum_{j}\left(2\rho-2\hat{\sigma}^{z}_j\rho\hat{\sigma}^z_j\right).
\end{eqnarray}
The first term involving a commutator describes coherent evolution due
to the Ising interaction, and the various terms having subscripts ``$\mathrm{ud}$'', ``$\mathrm{du}$'', and
``$\mathrm{el}$'' correspond respectively to spontaneous relaxation,
spontaneous excitation, and dephasing\footnote{The ``ud'' and ``du'' subscripts remind
us that the corresponding decoherence processes change a spin state
from ``up'' to ``down''  (or vice versa) along the $z$ axis, while the ``el'' subscript
reminds us that dephasing is ``elastic'' in the sense that it does not
change the spin projection along the $z$ axis.}.  Equation (\ref{meqformal}) has the formal solution $\rho(t)=\mathscr{U}(t)\rho(0)$, with
\begin{equation}
\mathscr{U}(t)=\exp\left[-t\left(i\mathscr{H}+\mathscr{L}_{\mathrm{ud}}+\mathscr{L}_{\mathrm{du}}+\mathscr{L}_{\mathrm{el}}\right)\right].
\end{equation}
The exponential of super-operators is meant to be understood via its
series expansion, in which the multiplication of two Lindblad
super-operators implies composition
$(\mathscr{L}_1\times\mathscr{L}_2)\left(\rho\right)=\mathscr{L}_1\left(\mathscr{L}_2\left(\rho\right)\right)$.
Our goal in what follows is to compute the time dependence of an
arbitrary operator $\hat{\mathcal{A}}$ at time $t$, given in the
Schr\"odinger picture by
\begin{equation}
\mathcal{A}(t)=\trace\left[\hat{\mathcal{A}}\rho(t)\right].
\end{equation}

\subsection{Dephasing ($T_2$ decoherence)}

An immediate simplification follows from the observation that
\begin{equation}
\left[\mathscr{L}_{\mathrm{el}},\mathscr{H}\right]=\left[\mathscr{L}_{\mathrm{el}},\mathscr{L}_{\mathrm{du}}\right]=\left[\mathscr{L}_{\mathrm{el}},\mathscr{L}_{\mathrm{ud}}\right]=0.
\end{equation}
That the last two commutators vanish is less obvious than the first,
but physically it has a very clear meaning: Spontaneous
relaxation/excitation on a site $j$ causes the $j$'th spin to have a well
defined value of $\sigma^z_j$, and thus to be unentangled with the rest of
the system.  Since the dephasing jump operator $\hat{\sigma}^z_j$ changes the relative phase
between the states $|\sigma^z_j=\pm1\rangle$, whether spontaneous
relaxation/excitation occurs before or after a dephasing event only
affects the \emph{sign} of the overall wave function,
which is irrelevant.  As a result, we can write
\begin{equation}
\mathscr{U}(t)=e^{-t\mathscr{L}_{\mathrm{el}}}e^{-t\left(i\mathscr{H}+\mathscr{L}_{\mathrm{ud}}+\mathscr{L}_{\mathrm{du}}\right)},
\end{equation}
and the time dependence of an arbitrary observable $\mathcal{A}(t)=\mathrm{Tr}\left[\rho(t)\hat{\mathcal{A}}\right]$ can
be written\footnote{Note that we apply the
operator $e^{-t\mathscr{L}_{\mathrm{el}}}$ to $\hat{\mathcal{A}}$
rather than $\rho(0)$.  This is justified by the identity
$\trace\left[\mathscr{L}_{\mathrm{el}}(\hat{\mathcal{O}}_1)\hat{\mathcal{O}}_2\right]=\trace\left[\hat{\mathcal{O}}_1
  \mathscr{L}_{\mathrm{el}}(\hat{\mathcal{O}}_2)\right]$, true for
arbitrary operators $\hat{\mathcal{O}}_{1}$ and $\hat{\mathcal{O}}_{2}$, which holds
because the jump operators $\hat{\sigma}^z_j$ are hermitian.}
\begin{equation}
\label{removeEL1}
\mathcal{A}(t)=\mathrm{Tr}\left[e^{-t\left(i\mathscr{H}+\mathscr{L}_{\mathrm{ud}}+\mathscr{L}_{\mathrm{du}}\right)}\rho(0)e^{-t\mathscr{L}_{\mathrm{el}}}\hat{\mathcal{A}}\right].
\end{equation}
Note that application of a superoperator does not commute
with operator products
($\mathscr{L}(\mathcal{O}_1)\mathcal{O}_2\neq\mathscr{L}(\mathcal{O}_1\mathcal{O}_2)$),
and we use the convention that a superoperator should act on the
operator immediately to its right.  The effect of the time evolution due to $\mathscr{L}_{\mathrm{el}}$
can be understood by considering its effect on the Pauli operators:
\begin{equation}
\label{removeEL2}
e^{-t\mathscr{L}_{\mathrm{el}}}\hat{\sigma}^{x,y}_{j}=e^{-\Gamma_{\mathrm{el}}t/2}\hat{\sigma}^{x,y}_j,~~~~~
e^{-t\mathscr{L}_{\mathrm{el}}}\hat{\sigma}^{z}_{j}=\hat{\sigma}^{z}_j.
\end{equation}
In light of equations (\ref{removeEL1}) and (\ref{removeEL2}), we are free to ignore the
dephasing terms in the master equation at the expense of attaching a
factor of $e^{-\Gamma_{\mathrm{el}}t/2}$ to every operator
$\sigma^{x,y}_j$ occurring inside an expectation value:
\begin{equation}
\label{removeEL3}
\mathrm{Tr}\left[\rho(t)\hat{\mathcal{A}}(\hat{\sigma}^x_j,\hat{\sigma}^y_j)\right]\rightarrow \mathrm{Tr}\left[\rho(t)\hat{\mathcal{A}}(e^{-\Gamma_{\mathrm{el}}t/2}\hat{\sigma}^x_j,e^{-\Gamma_{\mathrm{el}}t/2}\hat{\sigma}^y_j)\right],
\end{equation}
where $\rho(t)$ on the right-hand side evolves under the master
equation without the dephasing term.

\subsection{Spontaneous relaxation and excitation ($T_1$ decoherence)}

Because the effects of dephasing are fully included by
Eq. (\ref{removeEL3}), the remaining problem is to compute the time
dependence of operators whose expectation values are taken in a
density matrix evolving simultaneously under $\mathscr{H}$,
$\mathscr{L}_{\mathrm{ud}}$, and $\mathscr{L}_{\mathrm{du}}$:
\begin{equation}
\label{generalproblem}
\mathcal{A}(t)=\mathrm{Tr}\left[e^{-t\left(i\mathscr{H}+\mathscr{L}_{\mathrm{ud}}+\mathscr{L}_{\mathrm{du}}\right)}\rho(0)\hat{\mathcal{A}}\right].
\end{equation}
Formally the challenge of including the effects of
spontaneous relaxation and excitation is related to the nontrivial commutation
relation
\begin{equation}
\left[\mathscr{H},\mathscr{L}_{\mathrm{ud(du)}}\right]\neq0.
\end{equation}
Physically, the obstacle is that spontaneous relaxation and excitation
change the value of $\sigma^z$ for the spin which they affect, and
this change feeds back on the system through the Ising couplings.
From the results on coherent dynamics presented in Sec. \ref{sec1}, we know that time evolution
under $\mathscr{H}$ alone is tractable.  This suggests that we attempt
to solve Eq. (\ref{generalproblem}) by rewriting
\begin{equation}
i\mathscr{H}+\mathscr{L}_{\mathrm{ud}}+\mathscr{L}_{\mathrm{du}}=i\mathscr{H}_{\mathrm{eff}}-\mathscr{R},
\end{equation}
where $\mathscr{R}$ contains all and only terms that do not commute with $\mathscr{H}$, and then doing perturbation
theory in $\mathscr{R}$.  The above separation is accomplished by
defining an effective (non-Hermitian) Hamiltonian and its
corresponding superoperator
\begin{equation}
\label{hameffective}
\mathcal{H}_{\mathrm{eff}}=\mathcal{H}-i\gamma\sum_j\hat{\sigma}_j^z-i\mathcal{N}\frac{\Gamma_{\mathrm{r}}}{4}~~~~~\mathrm{and}~~~~~\mathscr{H}_{\mathrm{eff}}(\rho)=\mathcal{H}_{\mathrm{eff}}\rho-\rho\mathcal{H}_{\mathrm{eff}}^{\dagger},
\end{equation}
and the recycling term
\begin{equation}
\mathscr{R}(\rho)=\Gamma_{\mathrm{ud}}\sum_{j}\hat{\sigma}^{-}_j\rho\hat{\sigma}^{+}_{j}+\Gamma_{\mathrm{du}}\sum_{j}\hat{\sigma}^{+}_j\rho\hat{\sigma}^{-}_{j},
\end{equation}
in terms of which the time evolution operator is
$\mathscr{U}(t)=e^{-t(i\mathscr{H}_{\mathrm{eff}}-\mathscr{R})}$.
In Eq. (\ref{hameffective}) we have defined
$\gamma=\frac{1}{4}(\Gamma_{\mathrm{ud}}-\Gamma_{\mathrm{du}})$ and
$\Gamma_{\mathrm{r}}=\Gamma_{\mathrm{ud}}+\Gamma_{\mathrm{du}}$.
Defining $\mathscr{U}_0(t)=e^{-it\mathscr{H}_{eff}}$, we can now expand
the time evolution operator as a power series in $\mathscr{R}$ in
order to obtain the time-dependent expectation value $\mathcal{A}(t)$:
\begin{eqnarray}
\fl
&&\mathcal{A}(t)=\nonumber\\
\fl
&&\sum_{n}\!\int_0^t \!\!\!dt_{n}\dots\!\!\int_0^{t_2}\!\!\! \!dt_1\tr\left[\hat{\mathcal{A}}\mathscr{U}_0(t-t_{n})\mathscr{R}\mathscr{U}_0(t_{n}-t_{n-1})\dots\mathscr{U}_0(t_2-t_1)\mathscr{R}\mathscr{U}_0(t_1)\rho(0)\right]\!.
\label{A_of_t}
\end{eqnarray}
This expansion is the underlying object being evaluated when Monte
Carlo wave function methods \cite{QuantumTrajectories} (quantum trajectories) are used to approximate the density matrix.  In \ref{app_A}, we show in detail how the series in
Eq. (\ref{A_of_t}) leads to an expression for $\mathcal{A}(t)$ in terms of the closed-time path ordered correlation
functions obtained in Sec. \ref{sec1}, and the summation of that
series is carried out in \ref{integrals}.  Here we will simply summarize
the calculation, and explain in physical terms the essential structure
of the Hamiltonian and decoherence that allows the result to be cast in closed form.

\begin{figure}[h!]
\centering
\includegraphics[width=13cm]{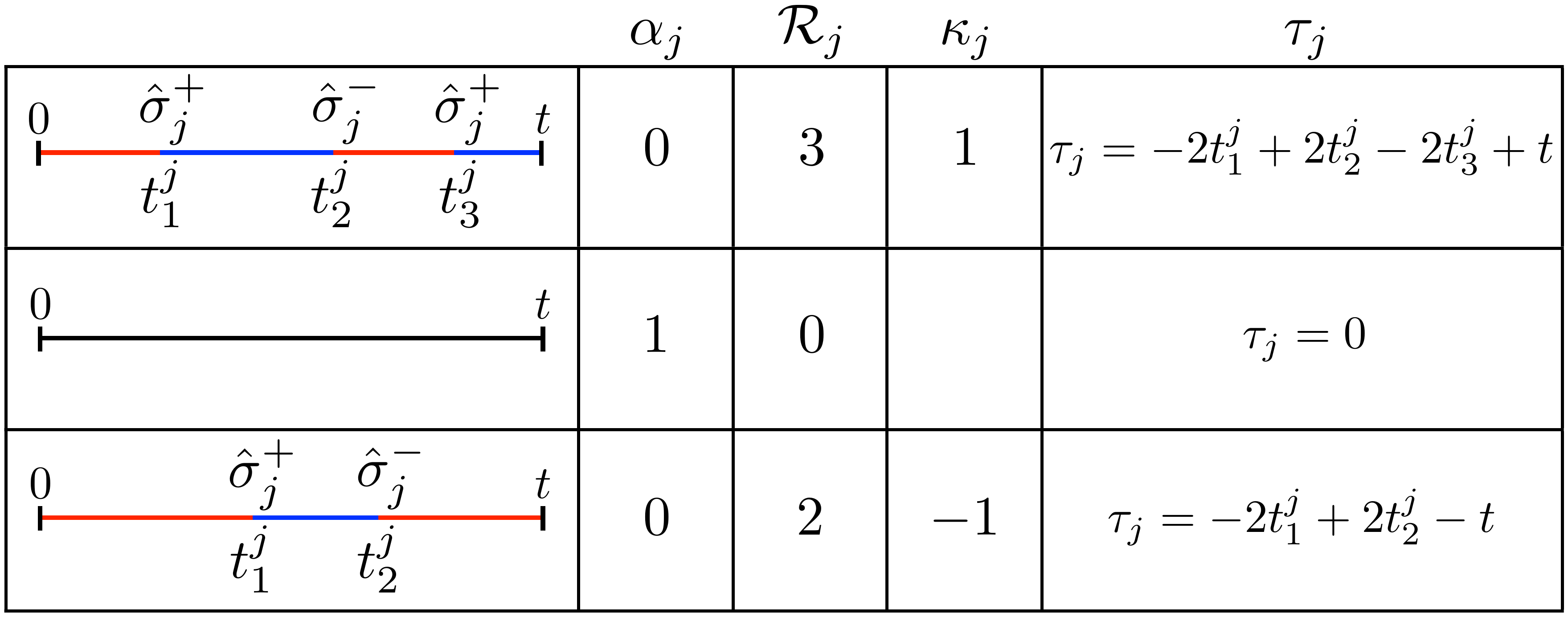}
\caption{A few examples of how spin-raising and spin-lowering
  operators belonging to the $j$'th site may occur along the forward
  time evolution, and the notation used to characterize these occurrences.}
\label{FigSymbdef}
\end{figure}

We begin by noting that when writing $\mathcal{A}(t)$ as a sum over closed-time path ordered correlation
functions, each operator inserted along the forward leg of the
time-contour is accompanied by its hermitian conjugate appearing at the same time along
the backward part of the time contour.  If we had explicitly included
an environment and attempted to trace over it (rather than starting
with a Markovian master equation), this feature of the problem would
emerge as a direct consequence of the Markov approximation.  As a result, it suffices to
describe any term in the series expansion by specifying the occurrence
of operators on the forward time contour.  To facilitate this description, we
introduce notation describing the occurrence of operators belonging to
a particular site $j$ (see Fig. \ref{FigSymbdef} for a summary).  We take $\mathcal{R}^{\pm}_j$ to
be the number of times the operator $\hat{\sigma}_j^{\pm}$ occurs
along the forward time path, $\{t_{1}^j,\dots,t_{\mathcal{R}_j}^j\}$ to be
the set of times at which jump operators are applied to site $j$, $\mathcal{R}_j=\mathcal{R}^{+}_j+\mathcal{R}^-_j$, $\kappa_j=\pm1$ depending on whether the
operator at the latest time along the forward path is
$\hat{\sigma}_j^{\pm}$, and
\begin{equation}
\tau_j=(1-\alpha_j)\int_{0}^{t}\sigma_j^{z}(t).
\end{equation}
We will also use bold symbols $\bm{\mathcal{R}}$, $\bm{\kappa}$, and $\bm{\tau}$ to
specify the complete set of these variables on all lattice sites.
Note that specifying $\mathcal{R}_j$ and $\kappa_j$ determines
both $\mathcal{R}^+_j$ and $\mathcal{R}_j^{-}$, since two consecutive (in the
closed-time-path-ordering) applications of an operator
$\hat{\sigma}_j^{\pm}$ gives zero.  Therefore, we will only
include the $\mathcal{R}_j$ and $\kappa_j$ as explicit arguments in
the correlation functions below.  These variables are sufficient to determine the value of any term in the series expansion
of $\mathcal{A}(t)$, so we do not need to keep track of the individual
times at which each jump operator is applied.

All nonzero terms in Eq. (\ref{A_of_t}) are captured by summing over
the $\mathcal{R}_j$ and $\kappa_j$, and integrating over the times
$t^j_1,\dots,t^j_{\mathcal{R}_j}$, denoted
\begin{equation}
\allsums=\prod_{j=1}^{\mathcal{N}}\allsums_{\!\!j}=\prod_{j=1}^{\mathcal{N}}\left(\delta_{\mathcal{R}_j,0}+\sum_{\kappa_j}\sum_{\mathcal{R}_j=1}^{\infty}\int_{0}^{t} dt_{\mathcal{R}_j}^{j}\dots\int_{0}^{t_2^j} dt_1^{j}\right).
\end{equation}
If $\hat{\mathcal{A}}$ can be written as a
product of operators $\hat{\sigma}^{b_j}_j$ ($b_j=\pm$) on sites $j$ contained in
a set $\eta$, then $\mathcal{A}(t)$ can be compactly expressed (see
\ref{app_A}) as
\begin{equation}
\label{formalsol}
\mathcal{A}(t)=\allsums \mathcal{P}(\bm{\mathcal{R}},\bm{\kappa},\bm{\tau})\mathcal{G}(\bm{s},\bm{\mathcal{R}},\bm{\tau}).
\end{equation}
The prefactor
\begin{equation}
\label{formalsol1}
\fl
\mathcal{P}(\bm{\mathcal{R}},\bm{\kappa},{\bm{\tau}})=\prod_j\mathcal{P}_j(\mathcal{R}_j,\kappa_j,\tau_j)=\prod_j\left(e^{-\Gamma_{\mathrm{r}}t/2}\left(\Gamma_{\mathrm{ud}}\right)^{\mathcal{R}^{-}_j}\left(\Gamma_{\mathrm{du}}\right)^{\mathcal{R}^{+}_j}e^{-2\gamma\tau_j}\right),
\end{equation}
is closely related to the probability that a series of jumps described
by the variables $\bm{\mathcal{R}}$, $\bm{\kappa}$, and $\bm{\tau}$
has occurred.  Defining the symbol $\bm{s}$ as the vector of quantities
$s_j=\varphi_j/t-2i\gamma$, the correlation functions
\begin{equation}
\label{formalsol2}
\fl
\mathcal{G}(\bm{s},\bm{\mathcal{R}},\bm{\tau})=\prod_j\mathcal{G}_j(s_j,\mathcal{R}_j,\tau_j)=\prod_{j}\left(e^{-i\vartheta_j(\tau_j)}\left[p_j+g_j^{+}(s_j t)\right]\right),
\end{equation}
are of the general form presented in Sec. \ref{sec1}.  Careful
bookkeeping reveals that the variables $\varphi$ and $\vartheta$ defined in Sec. \ref{sec1} are
given by
\begin{eqnarray}
\label{smallphi}
\varphi_j&=&2t\sum_{k\in\eta}b_kJ_{jk},\\
\vartheta(\bm{\tau})&=&\sum_{j}\vartheta_j(\tau_j)\\
\vartheta_j(\tau_j)&=&
\left\{\begin{array}{ll}
    0  & \mbox{if}~ j\in\eta;\\
-2\tau_j\sum_{k\in\eta}b_kJ_{jk} & \mbox{if}~ j\notin\eta.\end{array} \right.
\end{eqnarray}
Note that the $\mathcal{R}_j$ dependence in $\mathcal{G}_j$ is hidden in the
implicit dependence of $p_j$ and $g^{+}_j$ on $\alpha_j$ (which for
$j\notin\eta$ satisfies $\alpha_j=\delta_{\mathcal{R}_j,0}$).  When evaluating the sums and integrals in
Eq. (\ref{formalsol}), one must keep in mind that, whenever $j\in\eta$, terms with
$\mathcal{R}_j\neq 0$ vanish because they contain the consecutive
application of either $\sigma^{+}_j$ or $\sigma^{-}_j$.  Physically,
the vanishing of such terms reflects the lack of coherence for any spin
that has undergone even a single spontaneous spin flip.

The factorization of $\mathcal{P}$ into functions $\mathcal{P}_j$ of local site variables $\{\mathcal{R}_j,\kappa_j,\tau_j\}$ is a direct consequence of the
single particle nature of the anti-Hermitian part of
$\mathcal{H}_{\mathrm{eff}}$.  Physically, this factorization occurs because
the dissipation we are considering is uncorrelated from site to site (in contrast to the collective relaxation processes that arise, e.g., in
the context of Dicke superradience \cite{PhysRev.93.99}).  The
factorization of the correlation function $\mathcal{G}$ into functions $\mathcal{G}_j$
of local site variables $\{\mathcal{R}_j,\tau_j\}$ is a more surprising result, and depends crucially on the occurrence of jump
operators at the same times on the forward and backward time evolution
(\ref{app_A}).  The $\gamma$ appearing in the argument of $g_j^{+}(s_j
t=\varphi_j-2i\gamma
  t)$ affects the value of any term in Eq. (\ref{formalsol}) in which no jump
  operators are applied to site $j$ (such that $\alpha_j=1$ and therefore $g_j^+(s_j t)\neq 0$).
  It arises from the term $-i\gamma\hat{\sigma}^z$ in
  $\mathcal{H}_{\mathrm{eff}}$ [Eq. (\ref{hameffective})], and
  decreases the expectation value of $\hat{\sigma}^z_j$ for $\gamma>0$
  (when spontaneous relaxation outweighs spontaneous excitation).
  This effect is often referred to as \emph{null
    measurement state reduction}: Gaining knowledge that the spin on
  site $j$ has not spontaneously flipped affects the expected value when measuring
  $\hat{\sigma}^z_j$.

Because $\mathcal{G}$ and $\mathcal{P}$ both factorize, we need only
to evaluate the quantities
\begin{equation}
\label{PhiUn}
\Phi_{j}(s_j,t)=\sum\!\!\int_j\mathcal{P}_j(\mathcal{R}_j,\kappa_j,\tau_j)\mathcal{G}_j(s_j,\mathcal{R}_j,\tau_j)
\end{equation}
in terms of which
\begin{equation}
\label{sol1}
\mathcal{A}(t)=\prod_j\Phi_j(s_j,t).
\end{equation}
The explicit dependence on $s_j$ is included to remind the reader that, after the
sums and integrals (over $\mathcal{R}_j$, $\kappa_j$, and
$t_1^j\dots t_{\mathcal{R}_l}^{j}$) have been carried out, $s_j$ is the \emph{only}
  site-dependent quantity on which $\Phi_j(s_j,t)$ depends.  We evaluate these sums and integrals in \ref{integrals}, obtaining
\begin{equation}
\label{PhiEval}
\fl
\Phi_j(s_j,t)\!=\!\left\{ \begin{array}{*2{l}}
       \!\!\!   e^{-\Gamma_{\mathrm{r}}t/2}p_j &
       \mbox{if}~j\in\eta;\vspace{0.2cm}\\

       \!\!\!   e^{-\Gamma_{\mathrm{r}}/2}\left[\cos\!\left(\!t\sqrt{\!s_j^2-r}\,\right)+\left(\frac{\Gamma_{\mathrm{r}}}{2}+i\varphi_j\cos\theta_j\right)t\sinc\!\left(\!t\sqrt{\!s_j^2-r}\,\right)\!\right] & \mbox{if}~j\notin\eta,\end{array} \right.
\end{equation}
where $r=\Gamma_{\mathrm{ud}}\Gamma_{\mathrm{du}}$.
 
If $\hat{\mathcal{A}}$ also contains an operator $\hat{\sigma}_{l}^z(t)$ (with
$l\notin\eta$), we denote its expectation value by
$\mathcal{A}^z_l(t)$.  The insertion of $\hat{\sigma}_l^z$ must be
dealt with at the point of Eq. (\ref{PhiUn}).  Keeping in mind the
discussion surrounding Eq. (\ref{GF}), and remembering that $s_l=\varphi_l/t-2i\gamma$, we must
replace $\Phi_l(s_l,t)$ with
\begin{equation}
\label{PsiUn}
\Psi_l(s_l,t)=\sum\!\!\int_l\mathcal{P}_l(\mathcal{R}_l,\kappa_l,\tau_l)\left((1-\alpha_l)\kappa_l+\alpha_l\frac{i}{t}\frac{\partial}{\partial
s_l}\right)\mathcal{G}_l(s_l,\mathcal{R}_l,\tau_l)
\end{equation}
Therefore we have
\begin{equation}
\label{sol3}
\mathcal{A}_l^z(t)=\Psi_l(s_l,t)\prod_{j\neq l}\Phi_j(s_j,t),
\end{equation}
and in \ref{integrals} we find
\begin{equation}
\fl
\label{PsiEval}
\Psi_{l}(s_l,t)=e^{-\Gamma_{\mathrm{r}}/2}\left[\cos\!\left(\!t\sqrt{\!s_j^2-r}\,\right)+\left(is_l+2\gamma-\frac{\Gamma_{\mathrm{r}}}{2}\cos\theta_l\right)t\sinc\!\left(\!t\sqrt{\!s_j^2-r}\,\right)\!\right].
\end{equation}

\subsection{A simple application: under-damped to over-damped transitions}

\begin{figure}[t!]
\centering
\includegraphics[width=8cm]{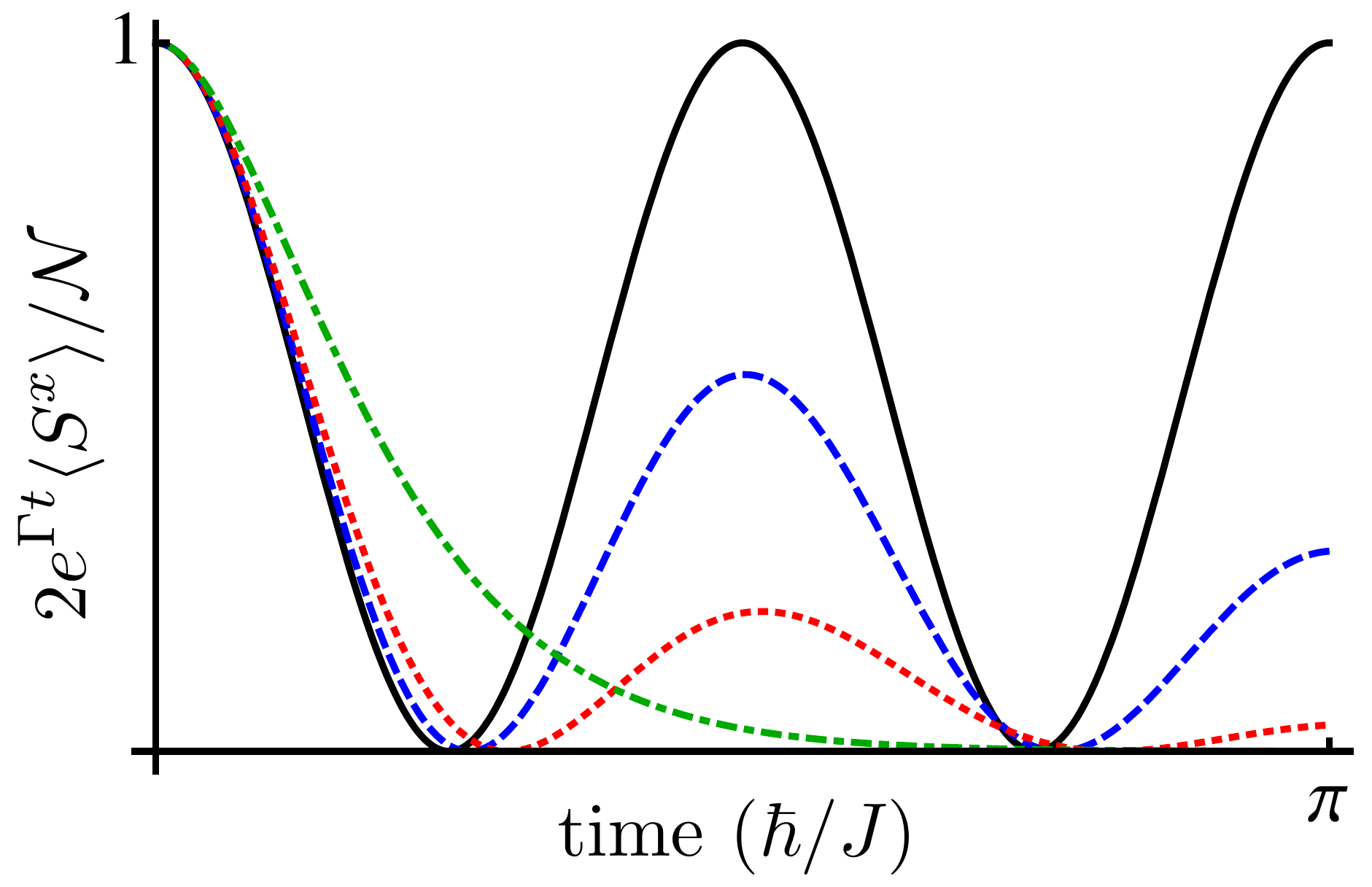}
\caption{Dynamics of the transverse spin length $S^x$ in a 1D nearest
  neighbor Ising model, with $\{\Gamma_{\mathrm{el}},\Gamma_{\mathrm{ud}},\Gamma_{\mathrm{du}}\}=\{0,\Gamma,\Gamma\}$, for several different values of
  $\Gamma$ between $0$
  (undamped) and $\Gamma_{\mathrm{c}}$ (critically damped): $\Gamma=0$
  (black solid line), $\Gamma=\Gamma_{\mathrm{c}}/8$ (blue dashed line), $\Gamma=\Gamma_{\mathrm{c}}/4$
  (red dotted line), and $\Gamma=\Gamma_{\mathrm{c}}$ (green dot-dashed line).}
\label{FigOscillations}
\end{figure}

These equations reveal that correlation functions will
generally undergo a qualitative transition in dynamics---from
over-damped to oscillatory---whenever the condition $s_l^2=r$ is
satisfied.  This behavior is the most clearly manifest when the
couplings $J_{ij}$ have a simple structure, such as nearest neighbor
or all-to-all.  For instance, for nearest-neighbor coupling in 1D,
assuming $\{\Gamma_{\mathrm{el}},\Gamma_{\mathrm{ud}},\Gamma_{\mathrm{du}}\}=\{0,\Gamma,\Gamma\}$, and choosing the
initial state to point along the $x-$axis, we find (ignoring boundary effects)
\begin{equation}
S^x(t)=\frac{\mathcal{N}}{2}e^{-3\Gamma t}\left[\cos\left(t\sqrt{4J^2-\Gamma^2}\right)+\Gamma t\sinc\left(t\sqrt{4J^2-\Gamma^2}\right)\right]^2,
\end{equation}
which becomes critically damped at $\Gamma_{\mathrm{c}}=2J$ (see
Fig. \ref{FigOscillations}).  It is interesting to note that this
solution is similar in structure to the damping of a classical harmonic
oscillator or a coherently driven two-level system (c.f. the weak-coupling limit of the spin boson problem \cite{caldeira,leggett}).  It is
important to contrast this behavior with that of a single spin
coupled to a Markovian bath, where the decoherence we consider only causes
a damped-to-oscillatory transition in the presence of a
\emph{transverse} magnetic field; the Hamiltonian
dynamics must be able to restore coherence in the basis for which the environment
induces a measurement.  In the present case, there is no transverse
field, and it is not \emph{a priori} obvious that such behavior should
emerge.  In fact, a simple mean-field estimate of the dynamics fails
to capture the oscillatory-to-damped transition.  Using a site-factorized ansatz for
the density matrix, $\rho=\bigotimes_j\rho_{j}$, it is straightforward to see that \cite{britton}
\begin{equation}
S^x_{\mathrm{MF}}(t)=\frac{\mathcal{N}}{2}e^{-\Gamma t}\cos(4 J\cos(\theta)t).
\end{equation}
Thus mean-field theory, which assumes an unentangled density matrix, \emph{always} predicts under-damped dynamics;
the transition to over-damped behavior captured by the exact solution depends crucially on
the competition between decoherence and entanglement.

\subsection{Qualitative insights into decoherence in interacting many-body systems\label{qual}}

In addition to providing an efficient way to compute arbitrary
observables for an open many-body system, the above calculation
provides significant insight into the interplay beween decoherence
and interactions.  To keep the notation as simple as possible, we will
focus on the case of nearest-neighbor coupling on a lattice with coordination number $z$, and calculate the time
dependence of the transverse spin length of a \emph{single} spin
$s^x(z,t)=\langle \sigma^x_j\rangle$ (for an infinite system it is
independent of $j$) for a state in which all spins are initially
polarized along the $x$-axis.  In the absence of decoherence but in
the presence of a longitudinal magnetic field of strength
$h$, application of results in Sec. \ref{sec1} gives
\begin{equation}
\label{coherentcase}
s^x_h(z,t)=\frac{1}{2}\mathrm{Re}\left[e^{-2iht}\cos^{z}(2Jt)\right].
\end{equation}

In the absence of a longitudinal field but including equal rates of
spontaneous relaxation/excitation
($\Gamma_{\mathrm{ud},\mathrm{du}}\equiv\Gamma$, $\gamma=0$),
we find instead [from Eqs. (\ref{sol1}) and (\ref{PhiEval})]
\begin{equation}
s^x(z,t)=e^{-(z+1)\Gamma_{\mathrm{r}}t/2}\left[\cos\left(t\sqrt{4J^2-\Gamma^2}\right)+\Gamma t\sinc\left(t\sqrt{4J^2-\Gamma^2}\right)\right]^{z}.
\end{equation}
However, it is instructive to temporarily hold off evaluating the sums and integrals implicit in
the $\Phi_j$, and work directly with Eqs. (\ref{PhiUn}) and (\ref{sol1}):
\begin{equation}
\label{uneval}
s^x(z,t)=\Phi_j(s_j,t)\prod_{k}^{\mathrm{NN}}\Phi_k(s_k,t)\prod_l^{\mathrm{D}}\Phi_l(s_l,t).
\end{equation}
In Eq. \ref{uneval} we have divided the product over lattice sites into three parts:
the $j$'th site, the nearest neighbor sites (product labeled
by NN) and the rest of the lattice (product labeled by D,
reminding us that these sites are Disconnected from site $j$).  First,
we observe that $\Phi_j(s_j,t)=\frac{1}{2}e^{-\Gamma_{\mathrm{r}}t/2}$.  Next we evaluate
$\prod^{\mathrm{D}}_l\Phi(s_l=0,t)=1$ [the $s_l=0$ because these sites
are disconnected from the $j$'th site, and hence $J_{jl}=0$, see
Eq. (\ref{smallphi})], which follows from a sum rule
$\sum\!\int_{l}\mathcal{P}_l(\mathcal{R}_l,\kappa_l,\tau_l)\mathcal{G}_l(0,\mathcal{R}_l,\tau_l)=1$
(derivable from $\tr\left[\rho(t)\right]=1$).  It remains only to
evaluate 
\begin{equation}
\sum^{\mathrm{NN}}\!\!\int\left(\prod^{NN}_{j}\mathcal{P}_j(\mathcal{R}_j,\kappa_j\tau_j)\mathcal{G}_j(s_j=2J,\mathcal{R}_j,\tau_j)\right),
\end{equation}
where we've adopted an abbreviated notation
$\sum^{\mathrm{NN}}\!\!\int=\prod_j^{\mathrm{NN}}\sum\!\int_j$
for the sums and integrals over the nearest-neighbor sites.  Utilizing\footnote{Note that in the final equality of
  Eq. (\ref{whyreal}), we have assumed that the left-hand side is
  real.  While this is not strictly true, the imaginary part of the
  left-hand side will vanish after the integral over $h$ is carried
  out below, so we make no mistake by ignoring it.}
\begin{equation}
\label{whyreal}
\prod_j^{\mathrm{NN}}\mathcal{G}_j(2J,\mathcal{R}_j,\tau_j)=2^{-\mathcal{R}}e^{-2iht}\cos(2Jt)^{z-\mathcal{R}}=\frac{2}{2^{\mathcal{R}}}s^x_h(z-\mathcal{R},t),
\end{equation}
where $\mathcal{R}=\sum_{j}^{\mathrm{NN}}\mathcal{R}_{j}$,
$\tau=\sum^{\mathrm{NN}}_j\tau_j$, and $h=J\tau/t$, and noting that $s_{h}^x(z-\mathcal{R},t)$ only depends on $\tau$ and
$\mathcal{R}$ (and not any other combinations of the local site
variables), we can then write
\begin{equation}
\label{suggestiveform}
s^x(z,t)=e^{-\Gamma_{\mathrm{r}}t/2}\sum_{\mathcal{R}=0}^{z}\int dh\; P(\mathcal{R},h) s^x_h(z-\mathcal{R},t).
\end{equation}
Here
\begin{equation}
P(\mathcal{R}',h')=\frac{t}{J}\sum^{\mathrm{NN}}\!\!\int\left(\prod^{\mathrm{NN}}_j\frac{\mathcal{P}_j(\mathcal{R}_j,\kappa_j,\tau_j)}{2^{\mathcal{R}_j}}\right)\delta(\tau-\tau')\delta_{\mathcal{R},\mathcal{R}'}
\end{equation}
is obtained by carrying out the sums and integrals while holding
$\mathcal{R}$ and $\tau$ fixed (the factor of $t/J$ arises when
changing variables from $\tau$ to $h$ in the remaining integral).

\begin{figure}[t!]
\centering
\includegraphics[width=14cm]{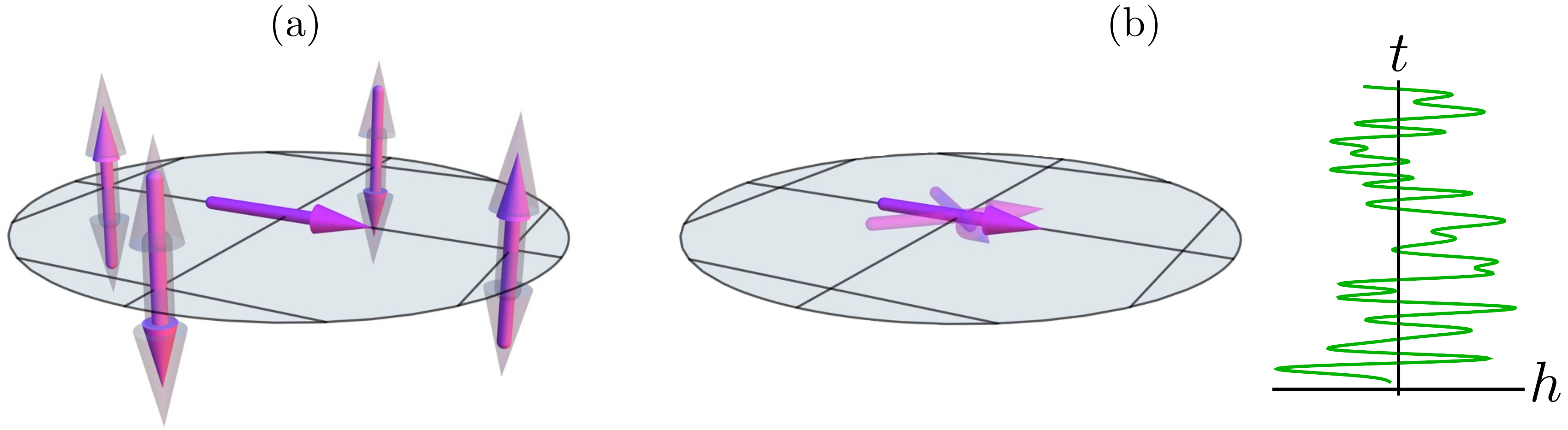}
\caption{The interplay between Ising interactions and spontaneous spin
flips induces a source of decoherence beyond the direct action of the
spin flips themselves.  In (a), the $j$'th (central) spin evolves due to the Ising couplings with its neighbors.
Spontaneous relaxation/excitation by its neighbors couples back (via the
Ising interactions) as a temporally fluctuating longitudinal field (b),
inducing a non-local dephasing process.}
\label{FigFlip}
\end{figure}

Equation (\ref{suggestiveform}) has a very suggestive form.  The factor of $e^{-\Gamma_{\mathrm{r}}t/2}$ out front is the
single-particle contribution of $T_{1}$ decoherence, and would be
present in the absence of interactions; it reflects the
probability that the $j$'th spin has not spontaneously flipped
before the time $t$ (a prerequisite for having any coherence along $x$).  As the $z$ nearest neighbors
evolve in time, they can undergo spin flips which cause them to
fluctuate in time between pointing along $+z$ and $-z$ [Fig. \ref{FigFlip}(a)].  Even once
flipped, they influence (via the Ising couplings) the $j$'th spin in
a manner formally equivalent to a longitudinal magnetic field of
strength $h=J(\tau/t)$ [Fig. \ref{FigFlip}(b)].
The quantity $P(\mathcal{R},h)$ describes the probability that $\mathcal{R}$ nearest neighbors have spontaneously
flipped, collectively contributing an effective magnetic field of
strength $h$ to the time evolution of the $j$'th spin.  The sum over
$\mathcal{R}$ and integral over $h$ then average the resulting dynamics for the
$j$'th spin, $s_h^x(z-\mathcal{R},t)$, over the possible behaviors of its neighbors.
For a given $\mathcal{R}$, the integral over $h$ reduces the Bloch vector length
by an amount depending on the width in $h$ of the distribution $P(\mathcal{R},h)$.
Physically, this integral captures the phase diffusion of the
$j$'th spin due to the stochastic temporal fluctuation of its immediate
environment (neighboring spins, as shown in Fig. \ref{FigFlip}).  Thus we see very clearly that the
interplay between spontaneous spin flips and coherent interactions leads
to an emergent source of dephasing: Flipping spins act
as fluctuating magnetic fields (mediated by the Ising interactions) on
other spins, even if these latter spins have not been directly affected by decoherence.

\section{One-axis twisting in an open system \label{sec3}}

The expressions in Eqs. (\ref{PhiEval}) and (\ref{PsiEval}) furnish a
complete description of correlation functions, and in special cases
afford descriptions of common experimental observables and
entanglement witnesses.  As a concrete example, we will use these
expressions to study the development and loss of entanglement in an
open-system version of the one-axis twisting (OAT) model.  It is
important to keep in mind, however, that most of the following results
can be generalized to take into account arbitrary Ising couplings.
Defining the collective spin operator $\hat{S}^z=\frac{1}{2}\sum_{j}\hat{\sigma}_j^z$, the OAT
Hamiltonian is given by 
\begin{equation}
\label{hamsat}
\mathcal{H}=2J(\hat{S}^z)^2,
\end{equation}
which is the $\zeta=0$ limit of our more general Ising model
[Eq. (\ref{modelham})].  For an initial state polarized along the
$x$-axis ($\theta=\pi/2$, $\phi=0$), it is well known \cite{ueda} that
the OAT Hamiltonian generates spin squeezed states at short times [the
squeezing is optimal at
$t_{\mathrm{s}}=\hbar\mathcal{N}^{-2/3}(3^{1/6}/2J)$], a fact that has
been exploited in a number of beautiful experiments
\cite{esteve,gross}.  In principle (i.e. in the absence of any
decoherence or other imperfections), these spin squeezed states allow
for precision metrology with a phase sensitivity that scales as $\mathcal{N}^{-5/6}$,
thus beating the $\mathcal{N}^{-1/2}$ scaling of the standard quantum
limit.  At time $t^{*}=\hbar\pi/4J$, the OAT Hamiltonian gives rise to a GHZ (or Schr\"odinger cat) state \cite{molmersorensen}, which
in principle affords Heisenberg limited ($\sim\mathcal{N}^{-1}$)
sensitivity in phase estimation.  In the following subsections, we
use the results of Sec. \ref{sec2} to extend calculate spin squeezing and characterize the metrological
utility of (and GHZ-type entanglement of) the state at t*, which would be the GHZ state in the absence of decoherence.

\subsection{Spin Squeezing}

Given the results in Sec. \ref{sec2}, analytic calculation of the squeezing parameter in the
presence of arbitrary decoherence rates $\Gamma_{\mathrm{el}}$,
$\Gamma_{\mathrm{ud}}$, and $\Gamma_{\mathrm{du}}$ is now
straightforward.
\begin{figure}[t!]
\centering
\includegraphics[width=14cm]{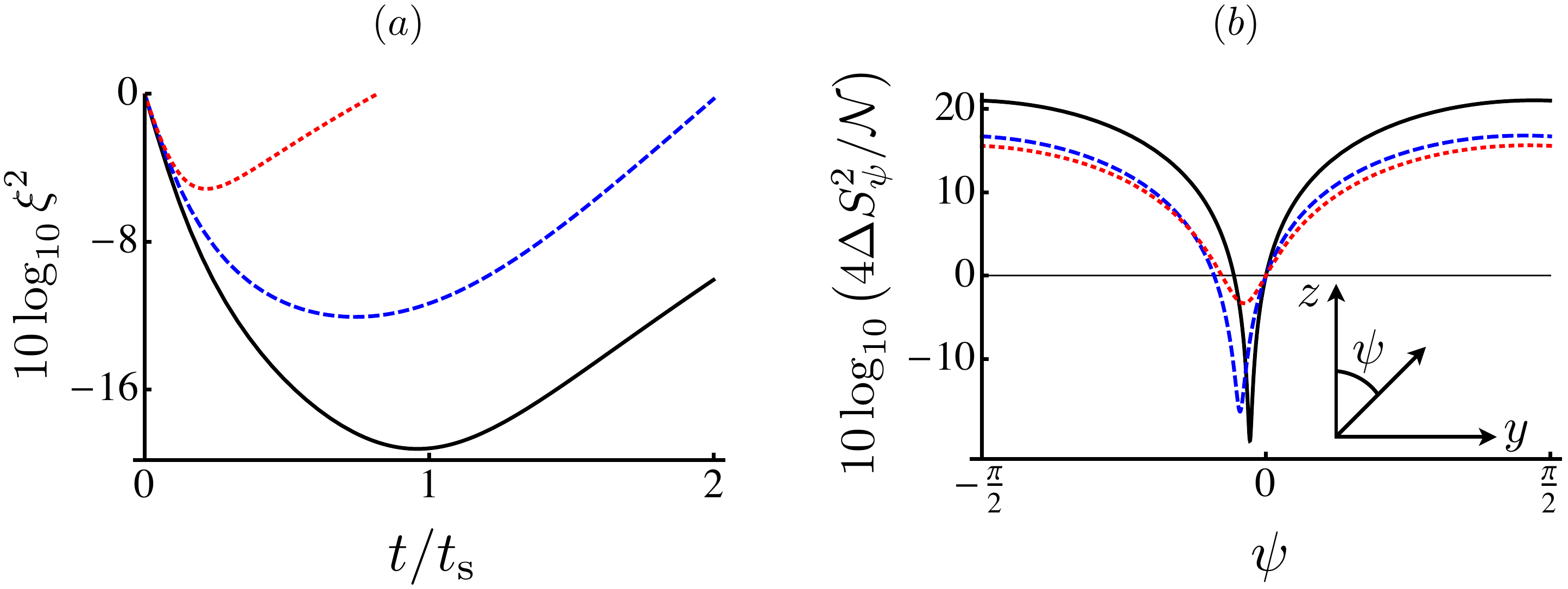}
\caption{Spin squeezing (dB) of $\mathcal{N}=10^{3}$ spins with no decoherence (black solid line), pure dephasing
  ($\{\Gamma_{\mathrm{el}},\Gamma_{\mathrm{ud}},\Gamma_{\mathrm{du}}\}=\{\Gamma,0,0\}$, blue dashed line), and equal amounts of
  spontaneous relaxation/excitation ($\{\Gamma_{\mathrm{el}},\Gamma_{\mathrm{ud}},\Gamma_{\mathrm{du}}\}=\{0,\Gamma/2,\Gamma/2\}$, red dotted
  line).  In (a) we show the optimal squeezing (optimized over angle
  $\psi$) as a function of time.  In (b) we plot the normalized
  variance, evaluated at time $t_{\mathrm{s}}$, as a function of the
  squeezing angle $\psi$.  Note that at the angle of
maximal squeezing ($\psi\approx0$), dephasing is much less detrimental than spontaneous
relaxation/excitation, whereas both have a similar effect at the angle
of maximal antisqueezing ($\psi\approx\pi/2$).  In all plots, we choose $\Gamma=1/t_{\mathrm{s}}$, with
  $t_{\mathrm{s}}=\hbar\mathcal{N}^{-2/3}(3^{1/6}/2J)$ being the time of optimal
  squeezing in the absence of decoherence.}
\label{FigSqueeze}
\end{figure}
As can be seen in Fig. \ref{FigSqueeze}(a), the effect of $T_2$ decoherence is
much less severe than that of $T_1$ decoherence.  One reason for this behavior
is that the minimum variance $\Delta S_{\mathrm{min}}^2$ occurs at an angle $\psi$ (in the
$yz$ plane) only slightly deviating from the $z$ axis.  Therefore dephasing,
which can be thought of as random rotations of the individual spins
around the $z$-axis, does not introduce much noise in the squeezed
quadrature (see Fig. \ref{FigSqueeze}b).  To the contrary, spontaneous
relaxation/excitation processes introduce noise directly into the
squeezed quadrature.

\subsection{Macroscopic superposition states \label{mss}}

In the absence of decoherence, the OAT Hamiltonian is known to give
rise to $\mathcal{N}$-spin GHZ states\footnote{Strictly speaking the
  form given only applies when the particle number $\mathcal{N}$ is
  even.  For odd $\mathcal{N}$ the GHZ state created looks similar,
  but is composed of states polarized along the $\pm y$ direction.}
\begin{equation}
\label{ghz}
 |\mathrm{GHZ}\rangle=\frac{|\Uparrow_x\rangle+i^{\mathcal{N}+1}|\Downarrow_x\rangle}{\sqrt{2}},
\end{equation}
at a time $t^{*}=\hbar\pi/4J$, where $|\Uparrow_x\rangle$ ($|\Downarrow_x\rangle$) denotes the state
where all spins point along the positive (negative) $x$-axis \cite{molmersorensen}.  These
entangled states afford Heisenberg-limited phase sensitivity \cite{bollingerCAT}, and
are a resource for certain types of fault-tolerant quantum
computation (\cite{leibfried} and references therein).  However, they are
also a canonical example of a fragile quantum state, and their
usefulness is easily destroyed by decoherence
\cite{PhysRevLett.79.3865}.  The effect of dephasing on the
production of GHZ state via one-axis twisting is well
understood \cite{PhysRevLett.79.3865,PhysRevA.77.052305}.  With the
results of Sec. \ref{sec2}, however, we can easily calculate the
effects of dephasing \emph{and} spontaneous relaxation/excitation on the
production of a GHZ state by one-axis twisting in a unified way.
In this section we explicitly compare the effects of dephasing to the
those of pure spontaneous relaxation ($\{\Gamma_{\mathrm{el}},\Gamma_{\mathrm{ud}},\Gamma_{\mathrm{du}}\}=\{0,\Gamma,0\}$).

\begin{figure}[t!]
\centering
\includegraphics[width=8cm]{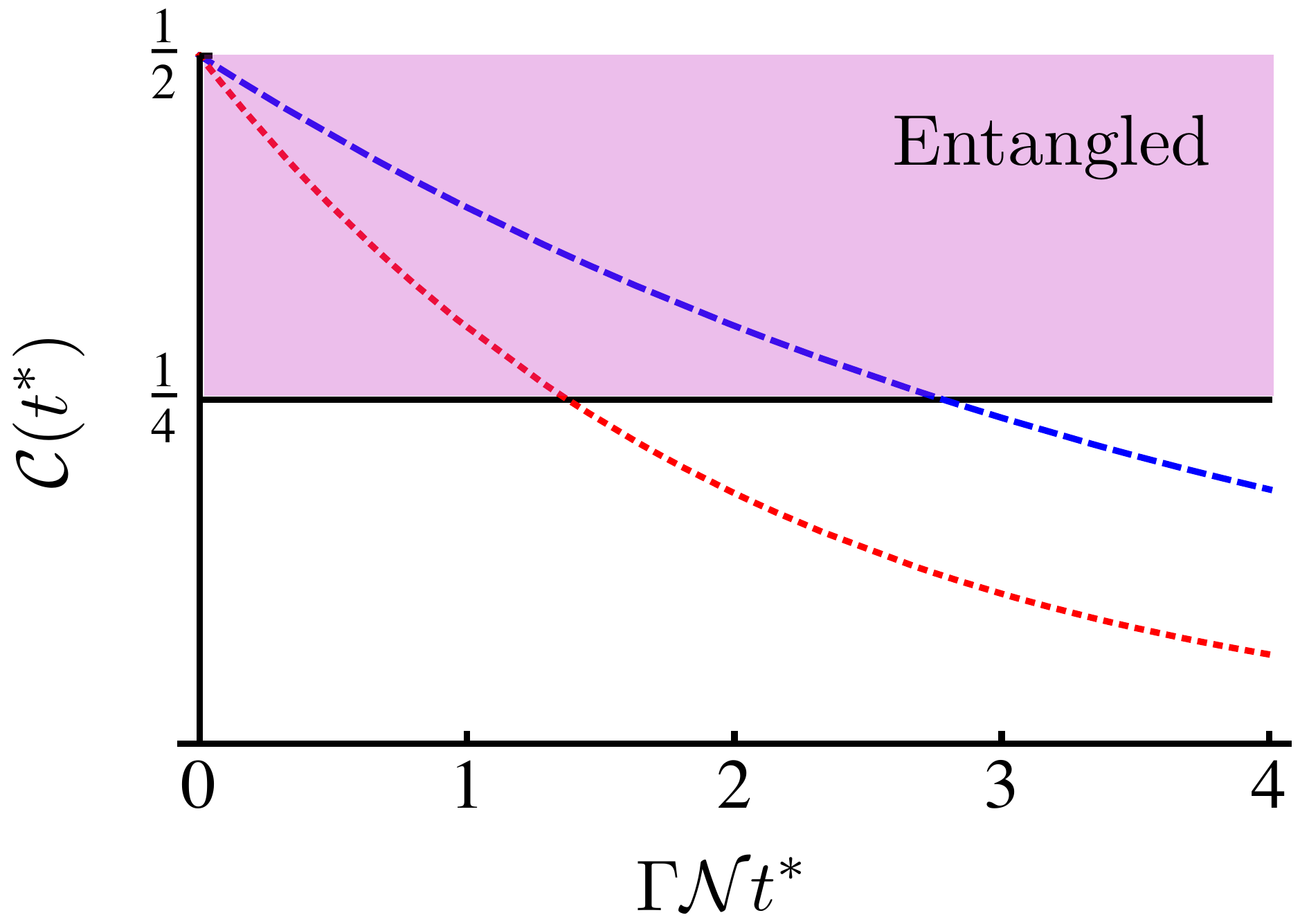}
\caption{Coherence of a GHZ state created in the presence of various
  types of decoherence:
  $\{\Gamma_{\mathrm{el}},\Gamma_{\mathrm{ud}},\Gamma_{\mathrm{du}}\}=\{\Gamma,0,0\}$
  (dephasing, dashed blue line) and $\{\Gamma_{\mathrm{el}},\Gamma_{\mathrm{ud}},\Gamma_{\mathrm{du}}\}=\{0,\Gamma,0\}$
  (spontaneous relaxation, dotted red line).  The region above the solid black line is
  guaranteed to have $\mathcal{N}$-particle GHZ type entanglement.}
\label{FigEW}
\end{figure}

We first characterize the GHZ state by its phase coherence, obtained from the
expectation value $\mathcal{C}=\trace\left[\rho(t)\hat{\mathcal{C}}\right]$ of the operator
\begin{eqnarray}
\hat{\mathcal{C}}&=&|\Uparrow_x\rangle\langle
  \Downarrow_x\!|\nonumber\\
&=&\frac{1}{2^{\mathcal{N}}}\prod_{j}(\hat{\sigma}_j^z+\hat{\sigma}_j^{+}-\hat{\sigma}_j^{-}).
\end{eqnarray}
The quantity $\mathcal{C}$ characterizes the extent to which the
superposition between the macroscopically distinct states
$|\Uparrow_x\rangle$ and $|\Downarrow_x\rangle$ is quantum mechanical (rather than a classical mixture).  Formally, $\mathcal{C}$ serves as a witness to $\mathcal{N}$-particle entanglement of the GHZ type, with
entanglement guarantied whenever $|\mathcal{C}|>1/4$ is satisfied \cite{sackett,leibfried}.  Application of the results in
Sec. \ref{sec2} yields
\begin{equation}
\mathcal{C}(t)=\frac{1}{2^{\mathcal{N}}}\sum_{m=1}^{\mathcal{N}}\sum_{n=0}^{m}{{\mathcal{N}}\choose{m}}{{m}\choose{n}}e^{-m\Gamma
t}\Psi\left(2J[2n-m]-2i\gamma,t\right)^{\mathcal{N}-m},
\end{equation}
and in the absence of decoherence one finds
$|\mathcal{C}(t^{*})|=1/2$.  As can be seen
in Fig. \ref{FigEW}, the effect of spontaneous relaxation on the
coherence $\mathcal{C}$ is comparable
to (but worse than) the effect of dephasing.  Both types of
decoherence cause a loss of phase coherence when $\Gamma\sim J/\mathcal{N}$, with the factor of
$\mathcal{N}$ responsible for the fragility of a GHZ state composed of a
large number of spins.

\begin{figure}[t!]
\centering
\includegraphics[width=8cm]{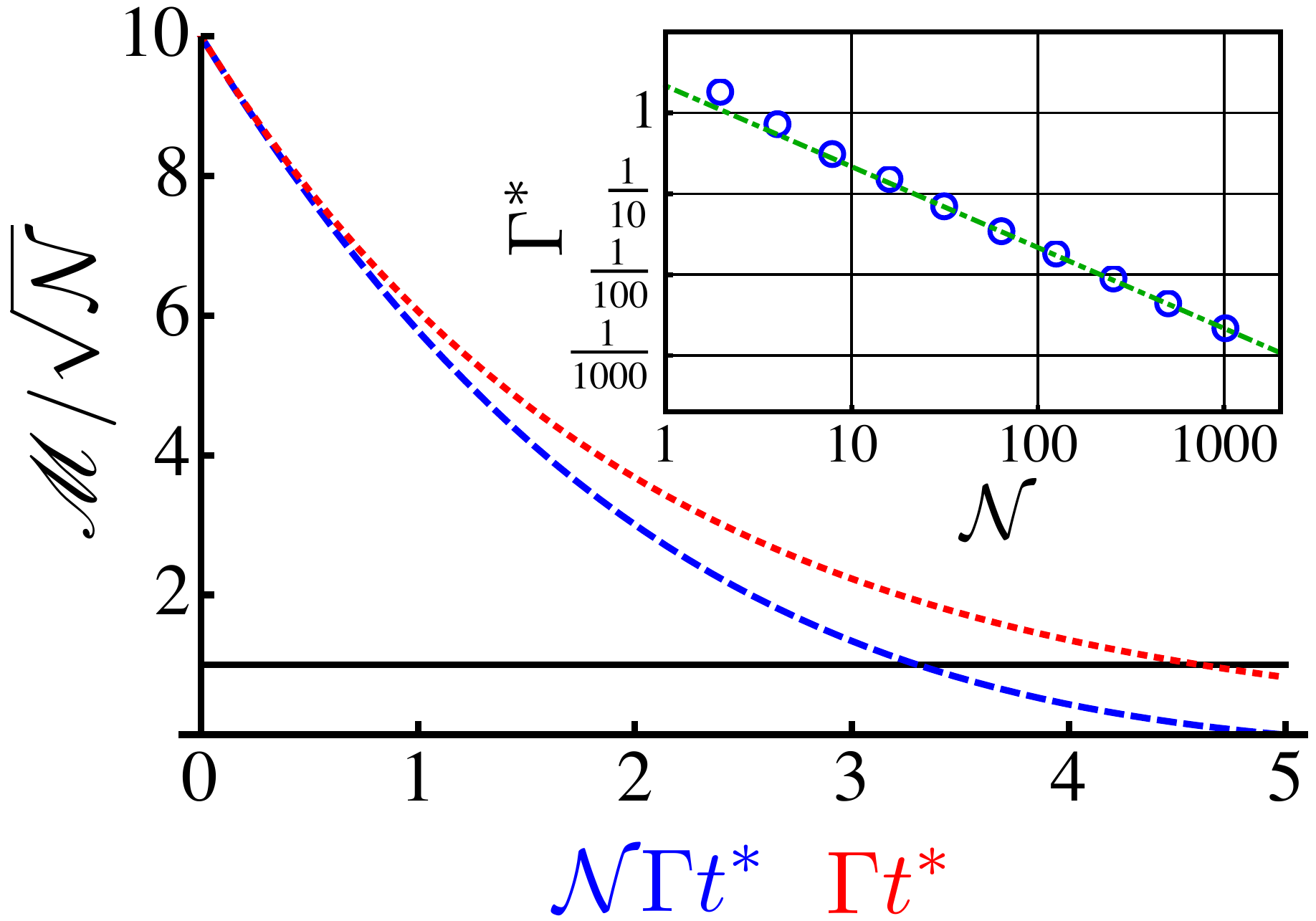}
\caption{Metrological gain over the standard quantum limit ($\mathscr{M}/\sqrt{\mathcal{N}}$) of $\mathcal{N}=100$ spins evolved under the OAT
  Hamiltonian to a time $t^{*}$ (where a GHZ state exists in the
  absence of decoherence).  We plot $\mathscr{M}$ as a function of decoherence
  rates for pure dephasing
  ($\{\Gamma_{\mathrm{el}},\Gamma_{\mathrm{ud}},\Gamma_{\mathrm{du}}\}=\{\Gamma,0,0\}$,
  red dotted line) and for spontaneous relaxation
  ($\{\Gamma_{\mathrm{el}},\Gamma_{\mathrm{ud}},\Gamma_{\mathrm{du}}\}=\{0,\Gamma,0\}$,
  blue  dashed line).  Note the two curves are produced by rescaling
  the decoherence rates in different ways (in order to show them in
  the same plot); if the scaling were the same for both plots, $\mathscr{M}$
  would decay \emph{much} more quickly for spontaneous relaxation than
  for dephasing.  Inset: Log-log plot of $\Gamma^{*}$, defined to be the
  decoherence rate for which $\mathscr{M}$ has decreased to $\mathcal{N}/e$, as a
  function of $\mathcal{N}$ (blue circles).  The green dot-dashed line
represents (up to a multiplicative constant) $1/\mathcal{N}$ scaling.}
\label{FigPrecision}
\end{figure}

We can also directly calculate the metrological usefulness of a
GHZ state prepared in the presence of spontaneous relaxation. The
favorable sensitivity of the state $\rho^{*}\equiv\rho(t^{*})$ to
rotations by angle $\Omega$ around the $x$-axis can be understood as the strong
dependence of the expectation value of the parity operator $\hat{\pi}=\prod_{j}\hat{\sigma}^z_j$,
\begin{eqnarray}
P(\Omega)&=&\tr\left[\rho^{*}(\Omega)\hat{\pi}\right]\\
&=&\tr\left[\rho^{*}\prod_{j}\left(\hat{\sigma}_j^z\cos\Omega-\hat{\sigma}_j^y\sin\Omega\right)\right]
\end{eqnarray}
on the angle $\Omega$ (where $\rho^{*}(\Omega)$ results from rotating
  $\rho^{*}$ about the $x$-axis by angle $\Omega$)
  \cite{bollingerCAT}.  The phase sensitivity of a GHZ state, denoted $\mathscr{M}$, is given (see Ref. \cite{PhysRevA.77.052305}) by
\begin{equation}
\label{Ggen}
\mathscr{M}=\left.\frac{\left|\partial
  P(\Omega)/\partial\Omega\right|}{\Delta P(\Omega)}\right|_{\Omega=0}\approx\sum_j|\trace[\rho^{*}\hat{\sigma}^{y}_j\prod_{k\neq j}\hat{\sigma}^{z}_k]|,
\end{equation}
where $\Delta P(\Omega)=1-P(\Omega)^2$ (taking into consideration that
$\hat{\pi}^2=1$) is the uncertainty of the operator
$\hat{\pi}$ calculated in the state $\rho^{*}$.  The approximation in
Eq. (\ref{Ggen}) is simply that $P(0)\approx
0$ and therefore $\Delta P(0)\approx1$.  This can be checked
explicitly by looking at the large $\mathcal{N}$ limit of
\begin{equation}
P(0)=\tr\left[\rho^{*}\hat{\pi}\right]=\left(\frac{2\gamma\left(e^{-\Gamma_{\mathrm{r}}t}-1\right)}{\Gamma_{\mathrm{r}}}\right)^{\mathcal{N}}.
\end{equation}
In the absence of decoherence, $\mathscr{M}=\mathcal{N}$ and the
Heisenberg limit of phase sensitivity is obtained.  By generalizing Eq. (\ref{sol3}) to the case where
$\hat{\sigma}_j^z$ is inserted on $\mathcal{N}-1$ of the sites, we
find that in the presence of decoherence
\begin{equation}
\mathscr{M}=\mathcal{N} e^{-\Gamma
  t/2}\;\mathrm{Im}\left[\Psi(2J-2i\gamma,t^{*})^{\mathcal{N}-1}\right].
\end{equation}
This result is plotted in Fig. \ref{FigPrecision} for different types
of decoherence.  The enhancement in $\mathscr{M}$ survives $T_1$
decoherence only if $\mathcal{N}\Gamma_{\mathrm{r}}
  t^{*}\lesssim1$ is satisfied (the scaling by $\mathcal{N}$ is shown
  in the inset).  This result should be contrasted with the effect of
  $T_2$ decoherence ($\{\Gamma_{\mathrm{el}},\Gamma_{\mathrm{ud}},\Gamma_{\mathrm{du}}\}=\{\Gamma,0,0\}$).
   In this case $\Psi(2J,t^{*})=1$, yielding
\begin{equation}
\mathscr{M}=\mathcal{N} e^{-\Gamma t/2},
\end{equation}
and hence the precision enhancement decays on a timescale that is
independent of $\mathcal{N}$ (consistent with results in Ref. \cite{PhysRevA.77.052305}).
In contrast, the entanglement witness $\mathcal{C}$ decays at an
$\mathcal{N}$-enhanced rate for either type of decoherence. 

\subsection{Removing the effects of decoherence via measurement-base feedback}

In Sec. \ref{mss}, we showed that spontaneous relaxation significantly degrades the
precision enhancement of a GHZ state unless $\Gamma\lesssim
J/\mathcal{N}$ is satisfied.  In this section, we will show that a
time-resolved record of spontaneous relaxation events provides sufficient information
to restore the phase enhancement under the much less stringent
constraint $\Gamma\lesssim J$.  In particular, we are imagining a
situation where spontaneous spin flips are accompanied by
the real spontaneous emission of a photon, such that they can be measured by photodetection.

For reasons that will become clear in what follows, we take our initial
state to have all spins pointing at an arbitrary angle $\theta$
(rather than $\theta=\pi/2$, as assumed in Sec. \ref{mss}).  Our goal is to
evaluate the expectation value of the operator $\sum_{j}\hat{\sigma}^y_j\prod_{l\neq
  j}\hat{\sigma}_l^z$ at time $t^{*}$, which was accomplished above by
appealing directly to Eqs. (\ref{PsiEval}) and (\ref{PhiEval}).
Pursuing a strategy similar to that employed in Sec. \ref{qual}, we
first observe that $n$ spins initialized at $\theta=\pi/2$, evolving in the absence of
decoherence but in the presence of a longitudinal field of strength
$h$, would yield the phase-sensitivity enhancement\footnote{Note that, for $h=0$ and $n$ odd,
  $\mathscr{M}(n,h)=0$.  This is because the GHZ state created when
  $n$ is odd is rotated from the GHZ states we consider by an angle of $\pi/2$ in
  the $xy$ plane.}
\begin{eqnarray}
\mathscr{M}(n,h)&=&n\left|\mathrm{Im}\left[e^{-2iht^{*}}(i\sin
  2Jt^{*})^{n-1}\right]\right|\\
&=&\left|\sin(2ht^{*}+\pi(n-1)/2)\right|.
\end{eqnarray}
Now we calculate $\mathscr{M}$ in the presence of decoherence, but we hold off evaluating the multiple sums and
integrals that yield the functions $\Psi_l(s_l,t)$ in closed form, and instead
obtain $\mathscr{M}$ by working directly with Eq. (\ref{formalsol})
\begin{equation}
\mathscr{M}=\allsums \mathcal{P}(\bm{\mathcal{R}},\bm{\kappa},\bm{\tau})\mathcal{G}(\bm{s},\bm{\mathcal{R}},\bm{\tau}).
\end{equation}
First, we evaluate $\mathcal{G}(\bm{s},\bm{R},\bm{\tau})$, obtaining
\begin{equation}
\mathcal{G}(\bm{s},\bm{\mathcal{R}},\bm{\tau})\!=\!(\mathcal{N}-\mathcal{R})\cos^{2\mathcal{R}}\!\left(\frac{\theta}{2}\right)\!\mathrm{Im}\!\left[e^{-2i J\tau}g^{-}(s=2Jt-2i\gamma t)^{\mathcal{N}-\mathcal{R}-1}\right]\!,
\end{equation}
which only depends on the $\mathcal{R}_j$ and $\tau_j$ only through
their sums $\mathcal{R}=\sum_j\mathcal{R}_j$ and
$h=(J/t)\sum_j\tau_j$.  With the judicious choice $\theta=\pi-2\tan^{-1}\left(e^{2\gamma t^{*}}\right)$,
we can rewrite
\begin{equation}
g^{-}(2Jt-2i\gamma t)=\frac{i\sin(2Jt)}{\cosh2\gamma t},
\end{equation}
and thus we obtain
\begin{equation}
\label{whereallsums}
\mathscr{M}=\allsums\mathcal{P}\left(\frac{\cos^{2\mathcal{R}}(\theta/2)}{\cosh^{N-1}(2\gamma
    t)}\right)\mathscr{M}(\mathcal{N}-\mathcal{R},J\tau/t).
\end{equation}
The initial value of $\theta$, which places the initial spins slightly
above the $xy$ plane, was carefully chosen so that
$g^{-}(2Jt-2i\gamma)\propto\sin(2Jt)$. This finite tipping angle is
required to precisely cancel the \emph{null measurement} effect, which causes the
$z$-projection of each spin to change in time due to the \emph{lack} of
emission of a photon, and thus ensures that at time $t^{*}$ the
unflipped spins are brought down into the $xy$ plane.  Finally, we can write
\begin{equation}
\mathscr{M}=\sum_{\mathcal{R}}\int dh\; P(\mathcal{R},h)\mathscr{M}(\mathcal{N}-\mathcal{R},h=J\tau/t),
\end{equation}
where $ P(\mathcal{R},h)$ is obtained by carrying out $\sum\!\int$ in
Eq. (\ref{whereallsums}) with
$\mathcal{R}$ and $\tau=h t/J$ held fixed.  As in section \ref{qual}, we can interpret $P(\mathcal{R},h)$ as the probability to have $\mathcal{R}$
flipped spins contributing an effective magnetic field $h$ to the
dynamics of the remaining spins.  For each particular value of
$\mathcal{R}$ and $h$, the function $\mathscr{M}(n,h)$ (where $n=\mathcal{N}-\mathcal{R}$)
yields the precision enhancement of a GHZ state produced from
$n$ spins in a longitudinal field of strength $h$.

If an experiment can record the times ($t_{1},\dots,t_{\mathcal{R}}$) at which photons are emitted,
thus gaining access to both $\mathcal{R}$ and
\begin{equation}
h=\frac{J}{t}\sum_{j=1}^{\mathcal{R}}(2t_j-t),
\end{equation}
then that experiment produces the \emph{conditional} density matrix $\rho(n,h)$,
corresponding to a GHZ state of $n$ spins produced by
one-axis twisting in the presence of the longitudinal field $h$.  The effect of
this field is simply to rotate the system around the $z$-axis by a
(shot-to-shot random) angle $\delta=(ht^{*}+\pi(n-1)/2)$.  However,
rotation by a random angle only causes decoherence if the value of
that angle is not known.  Because the experiment measures $\delta$ indirectly via the photon
emission record, it is possible to remove the effect of $h$ in any
experimental shot by applying the rotation operator $R(\delta)=\exp\left(-i S^z\delta\right)$ to the
conditional density matrix $\rho(n,h)$.  In this way we create
\begin{equation}
\rho(n,0)=R(\delta)\rho(n,h)R^{\dagger}(\delta),
\end{equation}
which is GHZ state of the form in Eq. (\ref{ghz}) containing
$n$ spins, and thus obtain a precision enhancement of $n$.  Because the
expected value of $n$ will decay only at the bare rate $\Gamma$,
there is no longer an enhancement by $\mathcal{N}$ in the decay of precision.
This measurement-based coherent feedback can also be applied in the
context of spin-squeezing, where once again it can vastly improve the
metrological usefulness of a state generated by one-axis twisting
in the presence of spontaneous relaxation.

Before concluding, we note that this feedback strategy could, in
principle, be applied to situations where both spontaneous relaxation
\emph{and} spontaneous excitation are present, and even when the
coupling constants $J_{ij}$ are not uniform.  However, in the former
case it is necessary to independently record the photon emission record
corresponding to excitation and relaxation processes, and in both cases it is necessary to obtain site-resolved (in addition to
time-resolved) information about the photon emissions.

\section{Conclusions and future directions \label{sec4}}

In this paper we have presented a comprehensive theoretical toolbox
for understanding far-from equilibrium dynamics in Ising models both
with and without decoherence.  The
underlying objects of interest are unequal-time
correlation functions, which are then used to compute spin squeezing,
dynamical response functions, entanglement
witnesses, and the effects of dephasing, spontaneous excitation, and
spontaneous relaxation on the system dynamics.  We believe these tools
will be of fundamental importance in understanding and optimizing a
diverse array of systems in which entanglement is engineered by Ising interactions.  In particular, these tools enable the
quantification of detrimental effects due to system-environment
coupling, even when the coupling does \emph{not} commute with the Ising interactions.

The ability to compute dynamics in any dimension and in the presence of
non-commuting noise is a particularly surprising result; it is well
known that the inclusion of non-commuting but \emph{coherent} linear
couplings admits solutions only in highly specialized geometries,
such as 1D nearest neighbor chains.  The key structures that allow our
solution for the open system to proceed are (1) the statistical
independence of decoherence processes on different sites and (2) a symmetry between the forward and
backward time evolution along a closed-time path, i.e. the Markov
approximation.  It would be interesting to understand to what extent
this simplification generalizes to other models where the
incorporation of decoherence would---at first sight---appear to be intractable.

We gratefully acknowledge Salvatore Manmana, Michael Kastner, Dominic
Meiser, Minghui Xu, and Murray Holland for helpful discussions.  This work was
supported by NIST, the NSF (PIF and PFC grants), AFOSR and ARO individual investigator awards, and the ARO
with funding from the DARPA-OLE program.  K. R. A. H. and M. F.-F. thank the NRC for
support.  This manuscript is the contribution
of NIST and is not subject to U.S. copyright.

\appendix

\section{\label{app_A}}

The main goal in this Appendix is to explicitly cast the series
expansion for arbitrary observables in terms of the time-ordered correlation
functions encountered in Sec. \ref{sec1} of the text, thus bridging
the gap between Eqs. (\ref{A_of_t}) and (\ref{formalsol}) of the text.  The summation of
the series is carried out later in \ref{integrals}.

Our starting point is the series expansion for the time-evolution
superoperator
\begin{equation}
\fl\mathscr{U}(t)=\sum_{n}\int dt_{n}\dots\int dt_1\mathscr{U}_0(t-t_{n})\mathscr{R}\mathscr{U}_0(t_{n}-t_{n-1})\dots\mathscr{U}_0(t_2-t_1)\mathscr{R}\mathscr{U}_0(t_1).
\end{equation}
This leads immediately to the expression for $\mathcal{A}(t)$ given in
the manuscript [Eq. (\ref{A_of_t}), reproduced here for convenience]:
\begin{eqnarray}
\fl
&&\mathcal{A}(t)=\nonumber\\
\fl
&&\sum_{n}\!\int_0^t \!\!\!dt_{n}\dots\!\!\int_0^{t_2}\!\!\! \!dt_1\tr\left[\hat{\mathcal{A}}\mathscr{U}_0(t-t_{n})\mathscr{R}\mathscr{U}_0(t_{n}-t_{n-1})\dots\mathscr{U}_0(t_2-t_1)\mathscr{R}\mathscr{U}_0(t_1)\rho(0)\right]\!.
\end{eqnarray}
In order to simplify notation in the following
equations, we define time dependent jump operators in the
Heisenberg picture of the effective Hamiltonian
\begin{equation}
\tilde{\mathcal{J}}(j,a,t)=\sqrt{\gamma^{a}}e^{it\mathcal{H}_{\mathrm{eff}}}\hat{\sigma}^{a}_{j}e^{-it\mathcal{H}_{\mathrm{eff}}},
\end{equation}
where $\gamma^{+(-)}=\Gamma_{\mathrm{du}(\mathrm{ud})}$.  Defining
\begin{equation}
\allsums\equiv \sum_{n=0}^{\infty}\sum_{j_1,\dots,j_n}\sum_{a_1,\dots,a_n}\int_{0}^{t} dt_{n}\dots\int_{0}^{t_2} dt_1,
\end{equation}
we can now express $\mathcal{A}(t)$ as
\begin{eqnarray}
\fl
&&\mathcal{A}(t)=\nonumber\\
\fl
&&\allsums\!\!\tr\!\left[\hat{\mathcal{A}}e^{-it\mathcal{H}^{\phantom\dagger}_{\mathrm{eff}}}\tilde{\mathcal{J}}(j_n,a_n,t_n)  \dots
\tilde{\mathcal{J}}(j_1,a_1,t_1)\rho(0)
\tilde{\mathcal{J}}^{\dagger}(j_1,a_1,t_1)\dots
\tilde{\mathcal{J}}^{\dagger}(j_n,a_n,t_n)
e^{it\mathcal{H}^{\dagger}_{\mathrm{eff}}}\!\right]\nonumber\\
\fl
&=&\allsums\left\langle \tilde{\mathcal{J}}^{\dagger}(j_1,a_1,t_1)\dots
\tilde{\mathcal{J}}^{\dagger}(j_n,a_n,t_n)\tilde{\mathcal{A}}(t)\tilde{\mathcal{J}}(j_n,a_n,t_n)  \dots
\tilde{\mathcal{J}}(j_1,a_1,t_1)\right\rangle.
\label{series}
\end{eqnarray}
In the above expression, the time dependence of the operator
$\tilde{\mathcal{A}}$ is defined as
\begin{equation}
\tilde{\mathcal{A}}(t)=e^{it\mathcal{H}^{\dagger}_{\mathrm{eff}}}\hat{\mathcal{A}}e^{-it\mathcal{H}^{\phantom\dagger}_{\mathrm{eff}}}
\end{equation}
which is \emph{distinct} from the time dependence assigned
to the operators $\hat{\sigma}_{j}^a$ in defining the $\tilde{\mathcal{J}}(j,a,t)$ (of
course this distinction vanishes when considering time evolution under a Hermitian Hamiltonian).  In the final line the trace has been removed (upon rearrangement, it
becomes a completeness identity), and the expectation value is in the
initial pure state $|\psi(0)\rangle$.

The correlation functions in Eq. (\ref{series}) are explicitly
closed-time path ordered, with the time ordering enforced in
the limits of integration.  The remaining task is to explicitly separate all time-dependence of the correlation
functions in Eq. (\ref{series}) due to the anti-Hermitian part of
$\mathcal{H}_{\mathrm{eff}}$, so that we can directly employ the results from
Sec. \ref{sec1}.  Because the anti-Hermitian part
of the effective Hamiltonian commutes with $\mathcal{H}$,
this is relatively straightforward.  As in the manuscript, we restrict
ourselves at this point to considering operators $\hat{\mathcal{A}}$
that can be written as products of spin-lowering and
spin-raising operators $\hat{\sigma}^{b_j}_j$ ($b_j=\pm1$) on sites
$j\in\eta$.  We can then write
\begin{equation}
\label{prefactor1}
\tilde{\mathcal{A}}(t)=\exp\left[-\frac{\mathcal{N}\Gamma_{\mathrm{r}}t}{2}\right]\exp\left[-2\gamma
  t\sum_{j\notin\eta}\hat{\sigma}_{j}^{z}\right]\hat{\mathcal{A}}(t),
\end{equation}
with $\hat{\mathcal{A}}(t)$ evolving in the Heisenberg picture of
$\mathcal{H}$ alone
\begin{equation}
\hat{\mathcal{A}}(t)=e^{it\mathcal{H}}\hat{\mathcal{A}}e^{-it\mathcal{H}}.
\end{equation}
Similarly, we can rewrite
\begin{equation}
\label{prefactor2}
\tilde{\mathcal{J}}(j,a,t)=e^{2a\gamma t}\sqrt{\gamma^{a}}\hat{\sigma}_j^a(t),
\end{equation}
with
\begin{equation}
\hat{\sigma}_j^a(t)=e^{i\mathcal{H}t}\hat{\sigma}_j^{a}e^{-i\mathcal{H}t}
\end{equation}
evolving in the Heisenberg picture of $\mathcal{H}$.

Replacing the operators in Eqs. (\ref{prefactor1}) and
(\ref{prefactor2}) back into Eq. (\ref{series}), we are essentially
ready to read off the results of a given term from the expressions for
correlation functions in Sec. \ref{sec1}.  However, anticipating that
the terms in the series will be site-factorizable, we pause here to
introduce some notation describing the occurrence of jump operators
belonging to a particular site.  Because jump operators always occur at the same time along the
 forward and backward evolution (and each operator on the backward
 path is the complex conjugate of a corresponding operator on the
 forward path), we only need to describe the jump operators applied
 during the forward time evolution.  First, we define $\mathcal{R}_{j}^{\pm}$ to be the total number of
jump operators of type $\hat{\sigma}_j^{\pm}$ applied to site $j$
along the forward time evolution, and
$\mathcal{R}_j=\mathcal{R}_j^{+}+\mathcal{R}_j^{-}$.  We also define
$\{t^j_{1},\dots,t^{j}_{\mathcal{R}_j}\}$ to be the set of times at
which those jump operators are applied to the site $j$, and
\begin{eqnarray}
\tau_j&=&(1-\alpha_j)\int_{0}^{t}\sigma_j^{z}(t)\\
&=&(1-\alpha_j)\kappa_j\left(t-2\sum_{n-1}^{\mathcal{R}_j}t_n^{j} (-1)^{\mathcal{R}_j-n}\right)
\end{eqnarray}
to be the total amount of time the $j$'th spin has spent
pointing up minus the time it has spend pointing down, again during
the forward evolution only (note that
$\int_{\mathcal{C}}\sigma^z_{j}(t)=0~\forall j\neq\eta$).  Finally we
use the bold symbols $\bm{\mathcal{R}},\bm{\kappa},\bm{\tau}$ to
represent vectors of the quantities $\mathcal{R}_j,\kappa_j$, and
$\tau_j$, respectively.

Now we can write $\mathcal{A}(t)$ as
\begin{eqnarray}
\fl
&&\mathcal{A}(t)=\nonumber\\
\fl
&&\allsums\!\!\mathcal{P}(\bm{\mathcal{R}},\bm{\kappa},\bm{\tau})\!\left\langle\!\!\hat{\sigma}^{-a_1}_{j_1}(t_1)\dots
\hat{\sigma}^{-a_n}_{j_n}(t_n)\hat{\mathcal{A}}(t)\hat{\sigma}^{a_n}_{j_n}(t_n)  \dots
\hat{\sigma}^{a_1}_{j_1}(t_1)\exp\!\!\left[\!2\gamma
  t\sum_{j\notin\eta}\alpha_j\hat{\sigma}_j^z\!\right]\!\right\rangle\!,
\label{squarebracket}
\end{eqnarray}
where
\begin{eqnarray}
\mathcal{P}(\bm{\mathcal{R}},\bm{\kappa},\bm{\tau})&=&\prod_{j}\mathcal{P}_j(\mathcal{R}_j,\kappa_j,\tau_j)\\
&=&\prod_{j}\left(e^{-\Gamma_{\mathrm{r}}t/2}\left(\Gamma_{\mathrm{ud}}\right)^{\mathcal{R}^{-}_j}\left(\Gamma_{\mathrm{du}}\right)^{\mathcal{R}^{+}_j}e^{-2\gamma\tau_j}\right).
\end{eqnarray}
The exponential of operators $\hat{\sigma}^z_j$ Eq. (\ref{squarebracket}) leads to the so-called \emph{null measurement state reduction}: it causes the $j^{\mathrm{th}}$ spin
to drift out of the $xy$ plane in the event that it has not spontaneously
relaxed ($\alpha_j=1$).
 
The expectation value in Eq. (\ref{squarebracket}) is now precisely of the form given in
Eq. (\ref{TOCF2}), with the exception that we must map
$\varphi_j\rightarrow s_j t\equiv\varphi_j-2i\gamma t$ in order to account for the
term in square brackets.  Thus we obtain
\begin{eqnarray}
\label{almostdone}
\fl
\mathcal{G}(\bm{s},\bm{\mathcal{R}},\bm{\tau})&\equiv&\left\langle\!\!\hat{\sigma}^{-a_1}_{j_1}(t_1)\dots
\hat{\sigma}^{-a_n}_{j_n}(t_n)\hat{\mathcal{A}}(t)\hat{\sigma}^{a_n}_{j_n}(t_n)  \dots
\hat{\sigma}^{a_1}_{j_1}(t_1)\exp\!\!\left[\!2\gamma
  t\sum_{j\notin\eta}\alpha_j\hat{\sigma}_j^z\!\right]\!\right\rangle\\
\fl
&=&e^{-i\vartheta(\bm{\tau})}\prod_{j}\left[p_j+g_j^{+}(s_j t)\right],
\end{eqnarray}
where $\bm{s}$ is a vector of quantities $s_j=\varphi_j/t-2i\gamma$.  Careful bookkeeping reveals that
\begin{eqnarray}
\varphi_j&=&2t\sum_{k\in\eta}b_kJ_{jk},\\
\vartheta(\bm{\tau})&=&\sum_{j}\vartheta_j(\tau_j)\\
\vartheta_j(\tau_j)&=&
\left\{\begin{array}{ll}
    0  & \mbox{if}~ j\in\eta;\\
-2\tau_j\sum_{k\in\eta}b_kJ_{jk} & \mbox{if}~ j\notin\eta,\end{array} \right.
\end{eqnarray}
which allows us to factorize
\begin{eqnarray}
\mathcal{G}(\bm{s},\bm{\mathcal{R}},\bm{\tau})&=&\prod_{j}\mathcal{G}_j(s_j,\mathcal{R}_j,\tau_j)\\
&=&\prod_{j}\left(e^{-i\vartheta_j(\tau_j)}\left[p_j+g_j^{+}(s_j t)\right]\right),
\end{eqnarray}
thus completing the derivation of Eqs. (\ref{formalsol}-\ref{formalsol2}) in the manuscript.

\section{\label{integrals}}

In Sec. \ref{sec2} we encountered the functions
\begin{eqnarray}
\fl
\Phi_j(s_j,t)&=&\sum\!\!\int_j\mathcal{P}_j(\mathcal{R}_j,\kappa_j,\tau_j)\mathcal{G}_j(s_j,\mathcal{R}_j,\tau_j)\\
\fl
\Psi_j(s_j,t)&=&\sum\!\!\int_j\mathcal{P}_j(\mathcal{R}_j,\kappa_j,\tau_j)\left((1-\alpha_j)\kappa_j+\alpha_j\frac{i}{t}\frac{\partial}{\partial
s_j}\right)\mathcal{G}_j(s_j,\mathcal{R}_j,\tau_j),
\end{eqnarray}
which we now show how to evaluate.  When $j\in\eta$, only the
$\mathcal{R}_j=0$ term in $\sum\!\int_j$ survives, and we obtain
\begin{equation}
\Phi_{j}(s_j,t)=e^{-\Gamma_{\mathrm{r}}t/2}p_j.
\end{equation}
The function $\Psi_j(s_j,t)$ is only defined for $j\notin\eta$, so
this case does not apply to it.  We next consider the case $j\notin\eta$, and
drop the site index $j$ (the derivation does not depend on the specific
site $j$, though the answer will depend on $s$, which can be reindexed
at the end).  We begin with $\Phi(s,t)$, which can be simplified as
\begin{eqnarray}
\fl
\Phi(s,t)&=&\left(\delta_{\mathcal{R},0}+\sum_{\kappa=\pm1}\sum_{\mathcal{R}=1}^{\infty}\int_{0}^{t}dt_{\mathcal{R}}\dots\int_{0}^{t_2}dt_1\right)\mathcal{P}(\mathcal{R},\kappa,\tau)\mathcal{G}(s,\mathcal{R},\tau)\\
\fl
&=&e^{-\Gamma_{\mathrm{r}}t/2}g^{+}(st)+\sum_{\kappa}\sum_{\mathcal{R}}\int_{0}^{t}dt_{\mathcal{R}}\dots\int_{0}^{t_2}dt_1\mathcal{P}(\mathcal{R},\kappa,\tau)pe^{-i\vartheta(\tau)}\\
\fl
&=&e^{-\Gamma_{\mathrm{r}}t/2}g^{+}(st)+e^{-\Gamma_{\mathrm{r}}t/2}\sum_{\kappa}\sum_{\mathcal{R}}\left(\Gamma_{\mathrm{ud}}\right)^{\mathcal{R}^{-}}\left(\Gamma_{\mathrm{du}}\right)^{\mathcal{R}^{+}}p\int_{0}^{t}dt_{\mathcal{R}}\dots\int_{0}^{t_2}dt_1\,e^{is\tau}\\
\fl
&\equiv&e^{-\Gamma_{\mathrm{r}}t/2}g^{+}(st)+e^{-\Gamma_{\mathrm{r}}t/2}\sum_{\kappa}\mathcal{X}(\kappa)
\end{eqnarray}
We will carry out the integrals first, for which
it is convenient to treat the cases $\mathcal{R}$ even and
$\mathcal{R}$ odd separately.  Remembering that
\begin{eqnarray}
\tau&=&(1-\alpha)\int_{0}^{t}\sigma^z(t)\nonumber\\
&=&\kappa\left[t-2t_{\mathcal{R}}+2t_{\mathcal{R}-1}\dots+(-1)^{\mathcal{R}}2t_1\right],
\end{eqnarray}
and defining a new index $\mu$ satisfying
\begin{equation}
\mu=\left\{\begin{array}{ccc}
\frac{\mathcal{R}-1}{2}&\mathrm{if}&\mathcal{R}~\mathrm{is}~\mathrm{odd}\\
&&\\
\frac{\mathcal{R}-2}{2}&\mathrm{if}&\mathcal{R}~\mathrm{is}~\mathrm{even,}
\end{array}\right.
\end{equation}
we obtain
\begin{equation}
\fl
\mathcal{X}(\kappa)=\left.\frac{t\left(1-\kappa\cos\theta\right)}{4}\sum_{\mu=0}^{\infty}\left(\Gamma_{\mathrm{r}}-4\kappa\gamma+r\frac{t-\kappa
    i\partial_z}{\mu+1}\right)\left(\frac{rt}{2z}\right)^{\mu}\frac{j_{\mu}(zt)}{\mu!}\right|_{z=s}.
\end{equation}
Here the $j_{\mu}$ are spherical Bessel functions, and the integral
has been carried out by changing variables to allow evaluation of all
but one integral while $\tau$ is held fixed, and then evaluating the
remaining integral over $\tau$ (this is where the Bessel functions
arise).  The remaining sum can be evaluated using relations obtained
from the generating function for spherical Bessel functions
\begin{eqnarray}
\mathcal{S}_{1}(x,y)&\equiv&\sum_{\mu=0}^{\infty}x^{\mu}\frac{j_{\mu}(y)}{(\mu+1)!}\\
&=&\sinc\sqrt{y^2-2xy}
\end{eqnarray}
and
\begin{eqnarray}
\mathcal{S}_{2}(x,y)&\equiv&\sum_{\mu=0}^{\infty}x^{\mu}\frac{j_{\mu}(y)}{(\mu)!}\\
&=&\frac{1}{xy}\cos\sqrt{y^2-2xy}-\frac{\cos y}{xy}.
\end{eqnarray}
In terms of these, we finally obtain
\begin{equation}
\fl
\mathcal{X}(\kappa)=\frac{t\left(1-\kappa\cos\theta\right)}{4}\left[\left(\Gamma_{\mathrm{r}}-4\kappa\gamma\right)\mathcal{S}_1\left(\frac{rt}{2s},st\right)+\left.r\left(t-i\kappa\partial_z\right)\mathcal{S}_2\left(\frac{rt}{2z},zt\right)\right|_{z=s}\right].
\end{equation}
Tedious but straightforward algebra leads to Eq. (\ref{PhiEval}) of Sec. \ref{sec2}.  Notice that we can just as easily evaluate the similar expression
\begin{eqnarray}
\fl
\Psi(s,t)&=&\!\left(\delta_{\mathcal{R},0}+\!\sum_{\kappa=\pm1}\sum_{\mathcal{R}=1}^{\infty}\int_{0}^{t}\!dt_{\mathcal{R}}\dots\int_{0}^{t_2}\!dt_1\right)\mathcal{P}(\mathcal{R},\kappa,\tau) \left((1-\alpha)\kappa+\alpha\frac{i}{t}\frac{\partial}{\partial
s}\right)\mathcal{G}(s,\mathcal{R},\tau)\nonumber\\
\fl
&=&e^{-\Gamma_{\mathrm{r}}t/2}g^{-}(st)+\sum_{\kappa}\sum_{\mathcal{R}}\int_{0}^{t}dt_{\mathcal{R}}\dots\int_{0}^{t_2}dt_1\mathcal{P}(\mathcal{R},\kappa,\tau)\kappa
pe^{-i\vartheta(\tau)}\\
\fl
&=&e^{-\Gamma_{\mathrm{r}}t/2}g^{-}(st)+e^{-\Gamma_{\mathrm{r}}t/2}\sum_{\kappa}
\mathcal{X}(\kappa)\kappa
\end{eqnarray}
giving rise to Eq. (\ref{PsiEval}) of Sec. \ref{sec2}.


\section*{References}

\begin{thebibliography}{10}

\bibitem{fossfeig1}
Foss-Feig M, Hazzard K~R A, Bollinger~J J, and Rey~A M.
\newblock Nonequilibrium dynamics of arbitrary-range ising models with
  decoherence: An exact analytic solution.
\newblock {\em Physical Review A}, 87:042101, 2013.

\bibitem{auerbach}
Auerbach A.
\newblock Interacting electrons and quantum magnetism.
\newblock 1994.

\bibitem{lacroix2011}
Lacroix C, Mendels P, and Mila F.
\newblock Introduction to frustrated magnetism.
\newblock 2011.

\bibitem{sachdev2008}
Sachdev S.
\newblock Quantum magnetism and criticality.
\newblock {\em Nat. Phys.}, 4:173, 2008.

\bibitem{hanson}
Hanson R, Kouwenhoven~L P, Petta~J R, Tarucha S, and Vandersypen L~M K.
\newblock Spins in few-electron quantum dots.
\newblock {\em Rev. Mod. Phys.}, 79:1217--1265, 2007.

\bibitem{prawer}
Prawer S and Greentree~A D.
\newblock Diamond for quantum computing.
\newblock {\em Science}, 320(5883):1601--1602, 2008.

\bibitem{mooij}
Chiorescu I, Nakamura Y, Harmans C J~P M, and Mooij~J E.
\newblock Coherent quantum dynamics of a superconducting flux qubit.
\newblock {\em Science}, 299(5614):1869--1871, 2003.

\bibitem{blochrmp}
Bloch I, Dalibard J, and Zwerger W.
\newblock Many-body physics with ultracold gases.
\newblock {\em Rev. Mod. Phys.}, 80:885--964, 2008.

\bibitem{poras}
Porras D and Cirac~J I.
\newblock Effective quantum spin systems with trapped ions.
\newblock {\em Phys. Rev. Lett.}, 92:207901, 2004.

\bibitem{sorensen}
Sorensen A, Duan L-M, Cirac~J I, and Zoller P.
\newblock Many-particle entanglement with bose–einstein condensates.
\newblock {\em Nature}, 409:63, 2001.

\bibitem{sachdevQPT}
Sachdev S.
\newblock Quantum phase transitions.
\newblock 1999.

\bibitem{PhysRevB.53.3304}
Schollw\"ock U, Jolic\oe{}ur Th, and Garel T.
\newblock Onset of incommensurability at the valence-bond-solid point in the
  \textit{S}=1 quantum spin chain.
\newblock {\em Phys. Rev. B}, 53:3304--3311, 1996.

\bibitem{kitaev}
Kitaev A.
\newblock Anyons in an exactly solved model and beyond.
\newblock {\em Ann. Phys.}, 321, 2006.

\bibitem{leggett}
Leggett~A J, Chakravarty S, Dorsey~A T, Fisher M~P A, Garg A, and Zwerger W.
\newblock Dynamics of the dissipative two-state system.
\newblock {\em Rev. Mod. Phys.}, 59:1--85, 1987.

\bibitem{PhysRevLett.78.167}
Antal T, R\'acz Z, and Sasv\'ari L.
\newblock Nonequilibrium steady state in a quantum system: One-dimensional
  transverse ising model with energy current.
\newblock {\em Phys. Rev. Lett.}, 78:167--170, 1997.

\bibitem{dallatorre}
Dalla Torre~E G, Demler E, Giamarchi T, and Altman E.
\newblock Quantum critical states and phase transitions in the presence of non
  equilibrium noise.
\newblock {\em Nature Physics}, 6:806--810, 2010.

\bibitem{PhysRevLett.97.236808}
Mitra A, Takei S, Kim~Y B, and Millis~A J.
\newblock Nonequilibrium quantum criticality in open electronic systems.
\newblock {\em Phys. Rev. Lett.}, 97:236808, 2006.

\bibitem{PhysRevLett.95.177201}
Feldman~D E.
\newblock Nonequilibrium quantum phase transition in itinerant electron
  systems.
\newblock {\em Phys. Rev. Lett.}, 95:177201, 2005.

\bibitem{diehl2}
Diehl S, Micheli A, Kantian A, Kraus B, B\"{u}chler~H P, and Zoller P.
\newblock Quantum states and phases in driven open quantum systems with cold
  atoms.
\newblock {\em Nat. Phys.}, 4:878, 2008.

\bibitem{prosen}
Prosen T.
\newblock Exact nonequilibrium steady state of a strongly driven open xxz
  chain.
\newblock {\em Phys. Rev. Lett.}, 107:137201, 2011.

\bibitem{kraus}
Kraus B, B\"uchler~H P, Diehl S, Kantian A, Micheli A, and Zoller P.
\newblock Preparation of entangled states by quantum markov processes.
\newblock {\em Phys. Rev. A}, 78:042307, 2008.

\bibitem{verstraete}
Verstraete F, Wolf~M M, and Cirac~J I.
\newblock Quantum computation and quantum-state engineering driven by
  dissipation.
\newblock {\em Nat. Phys.}, 5:633--636, 2009.

\bibitem{fossfeig}
Foss-Feig M, Daley~A J, Thompson~J K, and Rey~A M.
\newblock Steady-state many-body entanglement of hot reactive fermions.
\newblock {\em Phys. Rev. Lett.}, 109:230501, 2012.

\bibitem{barreiro}
Barreiro~J T, M\"uller M, Schindler P, Nigg D, Monz T, Chwalla M, Hennrich M,
  Roos~C F, Zoller P, and Blatt R.
\newblock An open-system quantum simulator with trapped ions.
\newblock {\em Nature}, 470:486, 2011.

\bibitem{PhysRevLett.107.080503}
Krauter H, Muschik~C A, Jensen K, Wasilewski W, Petersen~J M, Cirac~J I, and
  Polzik~E S.
\newblock Entanglement generated by dissipation and steady state entanglement
  of two macroscopic objects.
\newblock {\em Phys. Rev. Lett.}, 107:080503, 2011.

\bibitem{PhysRevLett.109.020403}
Olmos B, Lesanovsky I, and Garrahan~J P.
\newblock Facilitated spin models of dissipative quantum glasses.
\newblock {\em Phys. Rev. Lett.}, 109:020403, 2012.

\bibitem{PhysRevA.85.043620}
Ates C, Olmos B, Garrahan~J P, and Lesanovsky I.
\newblock Dynamical phases and intermittency of the dissipative quantum ising
  model.
\newblock {\em Phys. Rev. A}, 85:043620, 2012.

\bibitem{lee}
Lee~T E, H\"affner H, and Cross~M C.
\newblock Collective quantum jumps of rydberg atoms.
\newblock {\em Phys. Rev. Lett.}, 108:023602, 2012.

\bibitem{GopalakrishnanLee}
Gopalakrishnan S, Lee~T E, and Lukin~M D.
\newblock Unconventional magnetism via optical pumping of interacting spin
  systems.
\newblock {\em arXiv}, 1304.4959:[cond--mat.quant--gas], 2013.

\bibitem{lamacraft_2011}
Lamacraft A and Moore J.
\newblock Potential insights into nonequilibrium behavior from atomic physics
  (chapter from ultracold bosonic and fermionic gases).
\newblock 2012.

\bibitem{Trotzky18012008}
Trotzky S, Cheinet P, F\"olling S, Feld M, Schnorrberger U, Rey~A M,
  Polkovnikov A, Demler~E A, Lukin~M D, and Bloch I.
\newblock Time-resolved observation and control of superexchange interactions
  with ultracold atoms in optical lattices.
\newblock {\em Science}, 319(5861):295--299, 2008.

\bibitem{hazzard}
Hazzard K~R A, Manmana~S R, Foss-Feig M, and Rey~A M.
\newblock Far-from-equilibrium quantum magnetism with ultracold polar
  molecules.
\newblock {\em Phys. Rev. Lett.}, 110:075301, 2013.

\bibitem{britton}
Britton~J W, Sawyer~B C, Keith~A C, Wang C-C J, Freericks~J K, Uys H, Biercuk~M
  J, and Bollinger~J J.
\newblock Engineered two-dimensional ising interactions in a trapped-ion
  quantum simulator with hundreds of spins.
\newblock {\em Nature}, 484:489, 2012.

\bibitem{ueda}
Kitagawa M and Ueda M.
\newblock Squeezed spin states.
\newblock {\em Phys. Rev. A}, 47:5138--5143, 1993.

\bibitem{esteve}
Est\'eve A, Gross C, Weller A, Giovanazzi S, and Oberthaler~M K.
\newblock Squeezing and entanglement in a bose–einstein condensate.

\bibitem{gross}
Gross C, Zibold T, Nicklas E, Est\'eve J, and Oberthaler~M K.
\newblock Nonlinear atom interferometer surpasses classical precision limit.
\newblock {\em Nature}, 464:1165, 2010.

\bibitem{leibfried}
Leibfried D, Knill E, Seidelin S, Britton J, Blakestad~R B, Chiaverini J,
  Hume~D B, Itano~W M, Jost~J D, Langer C, Ozeri R, Reichle R, and Wineland~D
  J.
\newblock Creation of a six-atom `schr\"odinger cat' state.
\newblock {\em Nature}, 438:639--642, 2005.

\bibitem{monz}
Monz T, Schindler P, Barreiro~J T, Chwalla M, Nigg D, Coish~W A, Harlander M,
  H\"ansel W, Hennrich M, and Blatt R.
\newblock 14-qubit entanglement: Creation and coherence.
\newblock {\em Phys. Rev. Lett.}, 106:130506, 2011.

\bibitem{monroe1}
Kim K, Chang M-S, Korenblit S, Islam R, Edwards~E E, Freericks~J K, Lin G-D,
  Duan L-M, and Monroe C.
\newblock Quantum simulation of frustrated ising spins with trapped ions.
\newblock {\em Nature}, 465:590--593, 2010.

\bibitem{diehl1}
Diehl S, Rico E, Baranov~M A, and Zoller P.
\newblock Topology by dissipation in atomic quantum wires.
\newblock {\em Nat. Phys.}, 7:971, 2011.

\bibitem{krauter}
Krauter H, Muschik~C A, Jensen K, Wasilewski W, Petersen~J M, Cirac~J I, and
  Polzik~E S.
\newblock Entanglement generated by dissipation and steady state entanglement
  of two macroscopic objects.
\newblock {\em Phys. Rev. Lett.}, 107:080503, 2011.

\bibitem{monroe2}
Islam R, Edwards~E E, Kim K, Korenblit S, Noh C, Carmichael H, Lin G-D, Duan
  L-M, Wang~J C-C, Freericks~J K, and Monroe C.
\newblock Onset of a quantum phase transition with a trapped ion quantum
  simulator.
\newblock {\em Nature Comm.}, 2:377, 2011.

\bibitem{emch}
Emch~G G.
\newblock Non-markovian model for the approach to equilibrium.
\newblock {\em J. Math Phys.}, 7:1198, 1966.

\bibitem{radin}
Radin C.
\newblock Approach to equilibrium in a simple model.
\newblock {\em J. Math Phys.}, 11:2945, 1970.

\bibitem{kastner2}
Van Den~Worm M, Sawyer~B C, Bollinger~J J, and Kastner M.
\newblock Relaxation timescales and decay of correlations in a long-range
  interacting quantum simulator.
\newblock {\em arXiv}, 1209.3697:[quant--phys], 2012.

\bibitem{PhysRevLett.35.1792}
Sherrington D and Kirkpatrick S.
\newblock Solvable model of a spin-glass.
\newblock {\em Phys. Rev. Lett.}, 35:1792--1796, 1975.

\bibitem{PhysRevLett.23.17}
Griffiths~R B.
\newblock Nonanalytic behavior above the critical point in a random ising
  ferromagnet.
\newblock {\em Phys. Rev. Lett.}, 23:17--19, 1969.

\bibitem{PhysRevLett.54.1321}
Randeria M, Sethna~J P, and Palmer~R G.
\newblock Low-frequency relaxation in ising spin-glasses.
\newblock {\em Phys. Rev. Lett.}, 54:1321--1324, 1985.

\bibitem{PhysRevA.46.R6797}
Wineland~D J, Bollinger~J J, Itano~W M, Moore~F L, and Heinzen~D J.
\newblock Spin squeezing and reduced quantum noise in spectroscopy.
\newblock {\em Phys. Rev. A}, 46:R6797--R6800, 1992.

\bibitem{Campbell17042009}
Campbell~G K, Boyd~M M, Thomsen~J W, Martin~M J, Blatt S, Swallows~M D,
  Nicholson~T L, Fortier T, Oates~C W, Diddams~S A, Lemke~N D, Naidon P,
  Julienne P, Ye~J, and Ludlow~A D.
\newblock Probing interactions between ultracold fermions.
\newblock {\em Science}, 324(5925):360--363, 2009.

\bibitem{PhysRevLett.103.260402}
A.~M. Rey, A.~V. Gorshkov, and C.~Rubbo.
\newblock Many-body treatment of the collisional frequency shift in fermionic
  atoms.
\newblock {\em Phys. Rev. Lett.}, 103:260402, Dec 2009.

\bibitem{swallows}
Swallows~M D, Bishof M, Lin Y, Blatt S, Martin~M J, Rey~A M, and Ye~J.
\newblock Suppression of collisional shifts in a strongly interacting lattice
  clock.
\newblock 331(6020):1043--1046, 2011.

\bibitem{Gallagher1994}
Gallagher T.
\newblock Rydberg atoms.
\newblock {\em Cambridge University Press, Cambridge}, 1994.

\bibitem{uys}
Uys H, Biercuk~M J, VanDevender~A P, Ospelkaus C, Meiser D, Ozeri R, and
  Bollinger~J J.
\newblock Decoherence due to elastic rayleigh scattering.
\newblock {\em Phys. Rev. Lett.}, 105:200401, 2010.

\bibitem{QuantumTrajectories}
Plenio~M B and Knight~P L.
\newblock The quantum-jump approach to dissipative dynamics in quantum optics.
\newblock {\em Rev. Mod. Phys.}, 70:101--144, 1998.

\bibitem{gardiner}
Gardiner C and Zoller P.
\newblock Quantum noise.
\newblock 1991.

\bibitem{PhysRev.93.99}
Dicke~R H.
\newblock Coherence in spontaneous radiation processes.
\newblock {\em Phys. Rev.}, 93:99--110, 1954.

\bibitem{caldeira}
Caldeira~A O and Leggett~A J.
\newblock {\em Annals of Physics}, 149:374--456, 1983.

\bibitem{molmersorensen}
M\o{}lmer K and S\o{}rensen A.
\newblock Multiparticle entanglement of hot trapped ions.
\newblock {\em Phys. Rev. Lett.}, 82:1835--1838, 1999.

\bibitem{bollingerCAT}
Bollinger~J J, Itano~W M, Wineland~D J, and Heinzen~D J.
\newblock Optimal frequency measurements with maximally correlated states.
\newblock {\em Phys. Rev. A}, 54:R4649--R4652, 1996.

\bibitem{PhysRevLett.79.3865}
Huelga~S F, Macchiavello C, Pellizzari T, Ekert~A K, Plenio~M B, and Cirac~J I.
\newblock Improvement of frequency standards with quantum entanglement.
\newblock {\em Phys. Rev. Lett.}, 79:3865--3868, 1997.

\bibitem{PhysRevA.77.052305}
Rey~A M, Jiang L, Fleischhauer M, Demler E, and Lukin~M D.
\newblock Many-body protected entanglement generation in interacting spin
  systems.
\newblock {\em Phys. Rev. A}, 77:052305, 2008.

\bibitem{sackett}
Sackett~C A, Kielpinski D, King~B E, Langer C, Meyer V, Myatt~C J, Rowe M,
  Turchette~Q A, Itano~W M, Wineland~D J, and Monroe C.
\newblock Experimental entanglement of four particles.
\newblock {\em Nature}, 404:256--259, 2000.

\end{thebibliography}
\end{document}